\documentclass{article}
            \usepackage{graphicx}
            \usepackage{amsmath,amssymb}
            \begin{document}
            \title{Gauge theory extension to include number scaling by boson field:
             Effects on some aspects of physics and geometry}
            \author{Paul Benioff,\\
            Physics Division, Argonne National
            Laboratory,\\ Argonne, IL 60439, USA \\
            e-mail:pbenioff@anl.gov}
            \maketitle

            \begin{abstract}In gauge theories, separate vector spaces, $\bar{V}_{x}$, are assigned to each space time point $x$. Freedom of basis choice  is expressed by unitary  operators that relate matter field values in neighboring $\bar{V}_{x}$. Here gauge theories are extended by replacing the single underlying set of complex scalars, $\bar{C}$, with separate  sets,  $\bar{C}_{x}$, at each $x$, and including choice freedom of scaling for each $\bar{C}_{x}$. This scaling is based on the mathematical logical definition of mathematical systems as structures  satisfying a set of relevant axioms.  In gauge theory Lagrangians, number scaling shows  as a scalar boson field, $\Theta$, for which mass is optional and whose coupling to matter fields is very small.

            Freedom of number scaling in gauge theories is extended to a basic model where separate number
            structures of each type are assigned to each point of a manifold, $M$. Separate collections, $\bigcup_{x}$, of all types  of mathematical systems based on numbers, are assigned to each  $x$ of $M$. Mathematics available to an observer, $O_{x}$, at $x$ is that in $\bigcup_{x}$. The $\Theta$ field  induces scaling between structures in the different $\bigcup_{x}$. Effects  of $\Theta$ scaling on some aspects of physics and geometry are described. The lack of experimentally observed scaling means that $\Theta(z)$ is essentially constant for all points, $z$, in a region, $Z$, that can be occupied by us as observers. This restriction on $\Theta$ does not apply to points outside $Z$.

            The effects of $\Theta$ scaling on line elements, curve lengths, and distances between points,
            are examined. $O_{z}'s$ description, using the mathematics of $\bigcup_{z}$ in $Z$, of these elements at far away points, $x$, outside $Z$, includes  scaling  from $x$ to $z$. Integrals over curves include scaling factors inside the integrals. One example shows that the time dependence of $\Theta$ as $\Theta(t)$ can be such that mathematical, physical, and geometric quantities approach zero as $t$ approaches zero. This mimics the big bang in that distances between points approach zero. In the same sense, $\Theta$ scaling can also mimic inflation and the accelerated expansion of space as described by dark energy. Examples of black and white scaling holes are described in which $\Theta(x)$ is plus or minus infinity at a point $x_{0}.$

            \end{abstract}

            \section{Introduction}
            Gauge theories  are quite important to physics.  One example of this importance is their use as  the basis of the standard model.  Their development is based in the idea of freedom of choice of basis vectors at different space time points. This concept was introduced by Yang, Mills, \cite{YangM}.  for isospin space along with the requirement that physical interactions be independent of any choice.

            In gauge theories, these ideas are applied to other properties of systems besides isospin. Matter fields, $\psi(x),$ are described as taking values in separate vector spaces, $\bar{V}_{x}$, at each space time point, $x$ \cite{Montvay}. Unitary operators, $U_{y,x},$ as elements of a gauge group, connect $\bar{V}_{x}$ to $\bar{V}_{y}.$ They map vectors in $\bar{V}_{x}$ to vectors in $\bar{V}_{y}.$

            Mathematically, the definition of vector spaces includes  an underlying scalar field of real or complex numbers. In the usual setup, this is taken care of by using just one set of complex numbers, $\bar{C}$, as the common scalar field for the different vector spaces.  This setup raises the question regarding why one should use separate vector spaces for each space time point but just one complex number structure for all points. This is  relevant because complex numbers are part of the definition of vector spaces as used in gauge theories.

             This work continues earlier work \cite{BenQT} in the investigation of the expansion of the usual setup   by assigning separate complex number structures, $\bar{C}_{x},$ to each point $x.$ In this case $\bar{C}_{x}$ is the scalar number field for $\bar{V}_{x}.$   The different complex number structures can be related to one another by the use of parallel transform operators \cite{Mack}. These correspond to or define the notion of same number value between the different complex number structures.

             Restriction of the maps between the different number structure to parallel transform operators, corresponds in gauge theory to restricting the gauge group to the identity map.  There would be no freedom of choice of basis vectors among the different  vector spaces.   Here the freedom of choice of basis vectors in the vector spaces is extended to the underlying complex number structures by including a freedom of choice of scaling factors  that relate numbers in one structure to those in another.  This is achieved by expanding the parallel transform maps to include  space time dependent scaling factors.

             This extension was developed in earlier work \cite{BenQT,BenSPIE2}  by first expanding the gauge group $U(n)$ to $GL(1,c)\times SU(n).$ The real part of $GL(1,c)$ appears in gauge theory Lagrangians  as a scalar boson field, $\Theta(x),$ that interacts very weakly with matter fields.

               This description of gauge theories with separate vector spaces and complex number structures at each point $x$   was expanded by considering a basic model of physics and mathematics in which separate mathematical structures of many different types are associated with each space time point.  Included are the different types of numbers, vector spaces, algebras, etc. Any mathematical system type that is based  on numbers is included. It was also assumed that "mathematics is local" in that the mathematics available to an observer $O_{x}$ at $x$ is limited to the structures at $x.$

             The effect of scaling induced by the boson field $\Theta$ on some aspects of physics was investigated. It was seen that in a local region, including us as observers, scaling has not been observed experimentally. It follows that the effect of $\Theta $ must be below experimental error in the local region.  However these experimental restrictions do not apply to values of $\Theta$ at cosmological distances or for very large structures.

             Since the earlier work is used in the new material presented here, it  is summarized in the first Sections,  \ref{GT}-\ref{RT}, of this paper.   Some new material is also included. Section \ref{GT} summarizes the  expansion of gauge theories to include the freedom of number scaling.  Section \ref{BM} describes the basic model with separate number and other mathematical structures associated to each point of a space and time manifold. The effects of $\Theta$ induced scaling on quantum physics are summarized. New material is described  on the effect of scaling on the equations of motion of a classical system.

             Section \ref{RT} describes the restrictions on the space and time dependence of $\Theta(x)$  in a region, $Z$, in which it is possible for us as observers, to carry out experiments on systems.  Here  $Z$ is arbitrarily chosen to be a region of space of radius about 1 light year centered on the solar systems. The size of $Z$ is not important.  However it should be a very small fraction of the whole universe.

              Most of the new work  is in Sections \ref{ETG} and \ref{EETG}. Some effects of $\Theta$ induced scaling on the geometric properties of space and time are described. Section \ref{ETG} describes  the effects  on  line elements, curve lengths, and  distances between points.  It is seen that for local entities  at a point, such as the line element $dx^{2}$  at $x$, scaling arises in the transfer of the description of $dx^{2}$ at $x$, to any other point $z$, such as the location of an observer. Scaling between $x$ and $z$ is present because $dx^{2}$ is an infinitesimal number value in $\bar{R}_{x},$  but the representation of $dx^{2}$ at $z$ is an infinitesimal number value in $\bar{R}_{z}.$  This scaling can be removed by letting $z=x$  be  the reference point. For quantities such as curve lengths that are described by integrals over space and/or time, a space and/or time dependent scaling factor occurs inside the integral. This scaling cannot be removed by changing the reference point.

              Some examples of the effects of $\Theta$ induced scaling on geometric properties are discussed in Section \ref{EETG}. In one simple example, $\Theta(x)=\Theta(\vec{x},t)=\Theta(t)$ is assumed to depend on the time, $t$, only and not on space. Then the dependence of $d\Theta(t)/dt$ on time determines the time rate of change of  line elements, curve lengths and distances between points at all locations in the universe.  If $d\Theta(t)/dt>0$, then line elements, curve lengths, and distances all expand as time increases. If the expansion rate accelerates, then the change in these geometric properties has some similarity  to the accelerated expansion of space ascribed to dark energy \cite{Li}.   If $d\Theta(t)/dt<0$, then line elements, curve lengths, and distances all contract as time increases.

              Other examples illustrate the effects of singularities in the values of scaling.  In these $\Theta(r,\Omega,t)=\Theta(r)$ is assumed to be time independent and spherically symmetric about some point $x_{0}.$ $r$ is the radial unscaled distance from $x_{0}$ to some point $x.$ Let $z$ be a point on the extension of the  radius vector from $x_{0}$ to $x.$  If $\Theta(r)\rightarrow\infty$ as $r\rightarrow 0$, then the scaled distance from $z$ to $x$ increases to infinity as $x\rightarrow x_{0}.$  If $\Theta(r)\rightarrow -\infty$ as $r\rightarrow 0$, then the scaled distance from $z$ to $x$  approaches a finite limit or barrier in that it  is less than the unscaled distance from $z$ to $x_{0}$.

              These properties are shown in detail for  specific examples where $\Theta(r)=K/r$  and either $K>0$ or $K<0.$ The case $K>0$ is called a "scaling black hole" because the scaled distance from $z$ to $x$ increases without bound as $x$ approaches $x_{0}.$ The case $K<0$ is referred to as a "scaling white hole" because the scaled distance approaches a barrier as $x\rightarrow x_{0}.$

               Following a brief summary section, the final conclusion section emphasizes some important aspects of this work. Included are brief discussions of the notion of sameness and of the fact that scaling refers to all mathematical quantities,  independent of their possible representation of physical systems. The need to explore possible connections, if any, between $\Theta$ and other scalar fields discussed in physics is noted.

            \section{Gauge Theories}\label{GT}
            \subsection{Usual setup}
            As noted, the usual setup for gauge theories begins with the assignment of separate $n$ dimensional vector spaces, $\bar{V}_{x}$, to each space time point $x$ \cite{Montvay,Utiyama}.  Matter fields, $\psi(x)$, take values in $\bar{V}_{x}$ for each $x.$ Let $y=x+\hat{\nu}dx$ be a neighbor point of $x.$ Let $U_{y,x}:\bar{V}_{x}\rightarrow\bar{V}_{y}$ be a unitary operator in the gauge group $U(n)=U(1)\times SU(n).$

             $U_{y,x}$ is usually represented in terms of a phase, $\phi(x)$ and elements of the Lie algebra $su(n)$ \cite{Montvay,Cheng} as\begin{equation}\label{Uyx}U_{y,x}=e^{i\phi(x)}\exp(\sum_{\mu}\Gamma_{\mu} (x)dx^{\mu})\end{equation} where\begin{equation}\label{Gamma}\Gamma_{\mu}(x) =-ig\sum_{j}s^{j}_{\mu}(x)\tau_{j}.\end{equation} Here $s^{j}_{\mu}(x)$ is an $x$ dependent real number, $g$ is a coupling constant, and $\tau_{\mu}$ is a generator of $su(n).$

            There is a mathematical problem here. This representation of the action of $U_{y,x}$  on $\psi(x)$ is a vector in $\bar{V}_{x}.$ It is not a vector in $\bar{V}_{y}.$

            This problem can be easily fixed by factoring $U_{y,x}$ into two unitary operators as in \begin{equation}\label{UZW}U_{y,x}=Z_{y,x}W_{y,x}.\end{equation} In this case $U_{y,x}$ becomes a parallel transformation map from $V_{x}$ onto $\bar{V}_{y}$ \cite{Mack,LC,OR}. With this definition $U_{y,x}$ defines, or corresponds to, the notion of "sameness" between $\bar{V}_{x}$ and $\bar{V}_{y}.$ $U_{y,x}\psi(x)$ is the same vector in $\bar{V}_{y}$ as $\psi(x)$ is in $\bar{V}_{x},$ and $U_{x,y}\psi(y)$ is the same vector in $\bar{V}_{x}$ as $\psi(y)$ is in $\bar{V}_{y}.$

            As a map from $\bar{V}_{x}$ onto $\bar{V}_{x},$  the representation of $W_{y,x},$  with $W_{y,x}$ replacing $U_{y,x}$ in Eq, \ref{Uyx}, is valid.  The vector, $W_{y,x}U_{x,y}\psi(y),$ is defined to be the representation of $\psi(y)$ on $\bar{V}_{x}.$ The unitarity of $U_{x,y}$ and Eq. \ref{UZW} give \begin{equation}\label{Uxy}U_{x,y}=U_{y,x}^{\dag}=W_{y,x}^{\dag}Z_{y,x}^{\dag}\end{equation}and \begin{equation}\label{VyxUxy}W_{y,x}U_{x,y}\psi(y)=Z_{y,x}^{\dag} \psi(y)=Z_{x,y}\psi(y).\end{equation}As is the case for $U_{y,x},$ the unitary operator $Z_{y,x}:\bar{V}_{x}\rightarrow \bar{V}_{y}$ has no representation either as a matrix of numbers or in terms of Lie algebra elements.

             The distinction between $U_{y,x}$ and $W_{y,x}$ does not appear to be mentioned in the usual treatment of gauge theories.  The possible reason is that it makes no difference in the results.  However, this distinction is important for this work.
            \subsection{Expansion}\label{E}
\begin{figure}[h!]\begin{center}
            \rotatebox{270}{\resizebox{90pt}{90pt}{\includegraphics[50pt,300pt]
            [270pt,520pt]{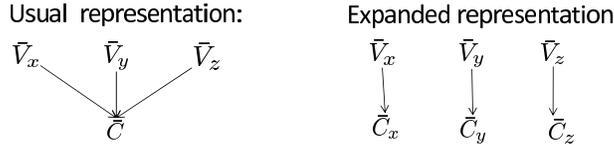}}}\end{center}
            \caption{Comparison of usual and expanded representations of scalar fields and vector spaces for gauge theory.}\label{ENS1}\end{figure}

            In the usual setup of gauge theories there is just one complex number field of scalars associated with each $\bar{V}_{x}$.  Here this is expanded by the association of separate complex number structures $\bar{C}_{x}$ with $\bar{V}_{x}$ for each $x.$ The couple $\bar{V}_{x},\bar{C}_{x},$ is assigned to each point $x$ rather than the couple $\bar{V}_{x},\bar{C}.$ This expansion is shown in Figure \ref{ENS1}.

            As was the case for the vector spaces, one introduces a parallel transform operator, \begin{equation}\label{Fyx}F_{y,x}:\bar{C}_{x}\rightarrow\bar{C}_{y}
            \end{equation}that maps $\bar{C}_{x}$ onto $\bar{C}_{y}.$  If $a_{x}$ is a number value in $\bar{C}_{x}$ then $a_{y}=F_{y,x}a_{x}$ is the same number value in $\bar{C}_{y}$ as $a_{x}$ is in $\bar{C}_{x}.$

             In order to proceed one needs a specific definition of  mathematical systems, such as vector spaces and complex, and other types of numbers. Here the  mathematical logic definition of mathematical systems   in general \cite{Barwise,Keisler} is used. A mathematical system of type $S$ is defined to be a structure, $\bar{S}=\{S,Op,Re,K\}.$ $S$ is a base set of mathematical elements,  $Op$ is a set of basic operations,  $Re$ is a set of basic relations, and  $K$ is a set of constants. The structure $\bar{S}$ is supposed to satisfy  a set of axioms for type $S$ systems.

            Relevant examples include real number structures, $\bar{R}=\{R,\pm,\times,\div,<,0,1\},$ that satisfy the axioms for a complete ordered field \cite{real}, complex number structures, $\bar{C}=\{C,\pm,\times,\div,0,1\}$ that satisfy the axioms for an algebraically complete field of characteristic $0$ \cite{complex}, and vector spaces $\bar{V}=\{V,\pm,\cdot,\psi\}$ that satisfy axioms for the type of vector space being considered. Here $\cdot$ denotes vector scalar multiplication and $\psi$ a general vector in the space.\footnote{A Hilbert space is a  complex, normed, inner product vector space that is complete in the norm defined from the inner product \cite{Kadison}. The representation as a structure is $\bar{H}=\{H,\pm,\cdot,\langle --\rangle,\psi\}.$}

            Use of these definitions of structures and the expansion to separate structures at each point, as in Fig. \ref{ENS1},  gives definitions of  $\bar{R}_{x},\bar{C}_{x},$ and $\bar{V}_{x}$ as
            \begin{equation}\label{RCVx}\begin{array}{c}\bar{R}_{x}=\{R_{x},\pm_{x},
            \times_{x},\div_{x},<_{x},0_{x},1_{x}\},\\\bar{C}_{x}=
            \{C_{x},\pm_{x},\times_{x},\div_{x},0_{x},1_{x}\},\\\bar{V}_{x}=
            \{V_{x},\pm_{x},\cdot_{x},\psi_{x}\}.\end{array}
            \end{equation}  Structures are distinguished from  their base sets by an overline, as in $\bar{C}_{x}$ vs. $C_{x}.$ Here $\bar{C}_{x}$ is the field of scalars for $\bar{V}_{x}.$

            As a parallel transformation \cite{Mack} of number structures, the map $F_{y,x}$ is an isomorphism from $\bar{C}_{x}$ to $\bar{C}_{y}$. Besides mapping the base set $C_{x}$ onto $C_{y},$ $F_{y,x}$ maps the operations  $\pm_{x},\times_{x},\div_{x}$ to $\pm_{y},\times_{y},\div_{y}.$ $F_{x,y}$ with reversed subscripts, is the inverse isomorphism. By extension $F_{y,x}$ and $F_{x,y}$ are also isomorphisms between $\bar{R}_{x}$ and $\bar{R}_{y}.$

            Scaling is accounted for by factoring $F_{y,x}$ into two isomorphic operators as in \begin{equation}\label{FXY}F_{y,x}\bar{C}_{x}=X_{y,x}Y_{y,x}\bar{C}_{x}=
            X_{y,x}\bar{C}^{r}_{x}.\end{equation}$Y_{y,x}$ maps $\bar{C}_{x}$ onto a scaled representation, $\bar{C}^{r}_{x},$ of $\bar{C}_{y}$ on $\bar{C}_{x},$ and $X_{y,x}$ maps $\bar{C}^{r}_{x}$ onto $\bar{C}_{y}.$ Here $r=r_{y,x}$ is a a real positive number in $\bar{C}_{x}.$

            The representation of $\bar{C}^{r}_{x}$ in terms of the elements, operations and number values in $\bar{C}_{x}$ is  \begin{equation}\label{Crx}\bar{C}^{r}_{x}=\{C_{x},\pm_{x}, \frac{\times_{x}}{r},r\div_{x},0_{x},r_{x}\}.\end{equation} An equivalent representation of  $\bar{C}^{r}_{x}$, as \begin{equation}\label{Crx1}\bar{C}^{r}_{x}=\{C_{x},\pm^{r}_{x}, \times^{r}_{x},\div^{r}_{x},0^{r}_{x},1^{r}_{x}\},\end{equation} shows, explicitly, the meaning of the operations  and constant values in Eq. \ref{Crx}.

            The scaling of the multiplication and division operations in Eq. \ref{Crx} is necessary so that $\bar{C}^{r}_{x}$ satisfies the complex number axioms if and only if $\bar{C}_{x}$ does.  This representation shows that $r_{x}=r\times_{x}1_{x},$ is the identity\footnote{This follows from the proof that $r_{x}$ is the multiplicative identity in $\bar{C}^{r}_{x}$ if and only if $1_{x}$ is the the multiplicative identity in $\bar{C}_{x}.$} in $\bar{C}_{x}^{r}$ even though it is not the identity in $\bar{C}_{x}.$

            An important consequence of the presence of $r$ factors with the scaled multiplication and division operations is that  products and quotients of terms in $\bar{C}^{r}_{x}$ end up with the same scaling factor as do single numbers.  For example, \begin{equation}\label{termr1}
            \frac{(a^{r}_{1})^{n_{1}}\times^{r}\cdots \times^{r}(a^{r}_{j})^{n_{j}}}
            {(b^{r}_{1})^{m_{1}}\times^{r}\cdots \times^{r}(b^{r}_{j})^{m_{k}}}\mbox{}^{r}\rightarrow r\frac{(a_{1})^{n_{1}}\times\cdots\times (a_{j})^{n_{j}}}{(b_{1})^{m_{1}}\times\cdots \times(b_{j})^{m_{k}}}.
            \end{equation} Here $a_{i}$ and $b_{i}$ are the same number values in $\bar{C}_{x}$ as $a^{r}_{i}$ and $b^{r}_{i}$ are in $\bar{C}^{r}_{x}.$ The $r$ factors associated with multiplication in the numerator cancel all but one factor associated with the number values.  This is canceled by the $r$ factor from the denominator. The remaining $r$ factor arises from $\div^{r}\rightarrow r\div.$

            It follows that for any analytic function $f(a),$ as the limit of a power series, \begin{equation}\label{analf}f^{r}(a^{r})\rightarrow rf(a).
            \end{equation}Here $f^{r}$ is the same function in $\bar{C}^{r}_{x}$ as $f$ is in $\bar{C}_{x}.$ One sees from this that equations are preserved under scaling from $\bar{C}^{r}_{x}$ to $\bar{C}_{x}.$  If $f^{r}(a^{r})=g^{r}(a^{r})$ in $\bar{C}^{r}_{x},$ then \begin{equation}\label{eqpres}f^{r}(a^{r}) =g^{r}(a^{r})\rightarrow rf(a)=rg(a)\rightarrow f(a)=g(a).\end{equation}This preservation of equations under scaling is important especially for theoretical predictions in physics that correspond to solutions of equations.

            These $r$ dependent representations emphasize the fact that the  elements of the base sets in the structures have no inherent number values outside of a number structure. They acquire values inside a structure only. These  values are determined by properties of the basic operations and relations in the structure. For example, the element of $C_{x}$ that has value $a^{r}_{x}$ in $\bar{C}^{r}_{x}$, Eq. \ref{Crx1},  has value $ra_{x}$ in $\bar{C}_{x}$. Here $a^{r}_{x}$ is the same number value in $\bar{C}^{r}_{x}$ as $a_{x}$ is in $\bar{C}_{x}.$

            This structure dependence of the values assigned to elements of $C_{x}$ is why the term "number values" is used instead of just "numbers". The elements of the base sets can be referred to as numbers.  However the value assigned to each base set element depends on the structure that contains the element.  More details are given in \cite{BenSPIE2}.

            Scaling also applies to the vector spaces. The scaled representation, $V^{r}_{x}$ of $\bar{V}_{y}$ on $\bar{V}_{x},$ is given by\footnote{\label{Hrx}The scaled representation, $\bar{H}^{r}_{x},$ of the Hilbert space, $\bar{H}_{y},$ on $\bar{H}_{x}$ is expressed by\\ $\bar{H}^{r}_{x}= \{H_{x},\pm_{x},\frac{\cdot_{x}}{r},\frac{\langle-,-\rangle_{x}}{r}, r\psi_{x}\}.$} \begin{equation}\label{Vrx}V^{r}_{x}=W_{y,x}U_{x,y}\bar{V}_{y}=W_{y,x}\bar{V}_{x}=
           \{V_{x},\pm_{x},\frac{\cdot_{x}}{r}, r\psi_{x}\}.\end{equation}  The presence of scaling means that the  definition of $W_{y,x}$ includes not only the Lie algebra representation of the gauge group, as in Eq. \ref{Uyx} with $W_{y,x}$ replacing $U_{y,x}$,  but also the effect of scaling. Here  the gauge group representation has been suppressed to simplify the expression.

           An equivalent representation of $\bar{V}^{r}_{x}$  as \begin{equation}\label{Vrx1}\bar{V}^{r}_{x}=\{V_{x},\pm^{r}_{x},\cdot^{r}_{x},\psi^{r}_{x}\}
            \end{equation} shows the meaning of the operations in $\bar{V}^{r}_{x}$ just as $\bar{C}^{r}_{x},$ Eq. \ref{Crx1}, shows the meaning of the operations in $\bar{C}^{r}_{x},$ Eq. \ref{Crx}. Here the $\bar{C}^{r}_{x},$ shown in Eq. \ref{Crx1} and in Eq. \ref{Crx}, are the respective scalar fields for the $\bar{V}^{r}_{x},$ shown in Eq. \ref{Vrx1} and in Eq. \ref{Vrx}. Also $\psi^{r}_{x}$ is the same vector in $\bar{V}^{r}_{x},$ Eq. \ref{Vrx1}, as $r\psi_{x}$ is in $\bar{V}^{r}_{x},$ Eq. \ref{Vrx} as $\psi_{x}$ is in $\bar{V}_{x}.$ More details on this, including support for inclusion of the factor $r$  in $r\psi_{x},$ is given in \cite{BenQT}.

            The dependence on $y$ arises through the definition of $r=r_{y,x}$ as the scaling factor of $\bar{C}_{y}$ relative to $\bar{C}_{x}.$ $r_{y,x}$ is a real number in $\bar{C}_{x}$ and $r_{x,y}$ is a real number in $\bar{C}_{y}.$ Note that the statement $r_{y,x}r_{x,y}=1$ makes no sense as multiplication is not defined between number structures, only within structures.  This is remedied by writing $r_{y,x}(r_{x,y})_{x}=r_{y,x}F_{x,y}(r_{x,y})=1_{x}.$  This is an equation in $\bar{C}_{x}.$

            The scaling factor, $r_{y,x},$  can be defined from a  new vector field,  $\vec{A}(x).$ If $y=x+\hat{\mu}dx$ is a neighbor point of $x$ then \begin{equation}\label{ryx}r_{y,x}=e^{\vec{A}(x) \cdot\hat{\mu}dx}.\end{equation}The association of separate complex number structures with each space time point means that for each $x,$ the exponent, and $r_{y,x},$ are real number values in $\bar{C}_{x}.$  If $y$ is distant from $x$, then \begin{equation}\label{ryxAd}r^{p}_{y,x} = \exp{(\int_{x}\vec{A}(z) \cdot\nabla_{z}pdz)}.\end{equation} Here $p$ is a path from $x$ to $y.$ The subscript $x$ means that the integral is evaluated in $\bar{C}_{x},$ and the superscript $p$ on $r_{y,x}$ indicates possible dependence on the path.

             In this work a simplification is used in that $\vec{A}$ is assumed to be the gradient of a scalar field $\Theta $ as in \begin{equation}\label{Ath}\vec{A}(x)=\nabla_{x}\Theta.\end{equation}Eqs. \ref{ryx} and \ref{ryxAd} become  \begin{equation}\label{ryxT}r_{y,x}= e^{\nabla_{x}\Theta\cdot\hat{\mu}dx}\end{equation} and \begin{equation}\label{ryxdT}r_{y,x}=e^{\Theta(y)_{x}-\Theta(x)}.
            \end{equation} The subscript $x$ on $\Theta(y)$ indicates that $\Theta(y)_{x}$ is the same value in $\bar{C}_{x}$ as $\Theta(y)$ is in $\bar{C}_{y}.$

            The advantage of this simplification is that $r^{p}_{y,x}$ is independent of the path $p.$ This occurs because the gradient theorem \cite{Gradient} allows replacement of the integral in Eq. \ref{ryxAd} by the endpoints.

             For gauge theories, the presence of the scaling factor  results in an expansion of the gauge group by replacing the $U(1)$ factor by $GL(1,c).$  If $\bar{V}_{x}$ is $n$ dimensional, the  gauge group, $U(n),$ is replaced by $GL(1,c)\times SU(n)$ \cite{Utiyama}.  The real component of $GL(1,c)$ is due to the $\Theta$ field.

             The covariant derivative for matter fields with the scaling factor included is
              \begin{equation}\label{Dmupsi}D_{\mu}\psi=\frac{r_{x+dx^{\mu},x}
              V(x+dx^{\mu},x)\psi(x+dx^{\mu})_{x}-\psi(x)}{dx^{\mu}}.\end{equation} Here $V(x+dx^{\mu},x)$ is an element of $U(n)$, and $r_{x+dx^{\mu},x}$ is given by  Eq. \ref{ryxT} with $\nabla_{x}$ replaced by the $\mu$ component, $\partial_{\mu,x}.$ The term, $r_{x+dx^{\mu},x}
              V(x+dx^{\mu},x)\psi(x+dx^{\mu})_{x},$ is obtained by noting that \begin{equation}\label{Wxdx}
              W_{x+dx^{\mu},x}U_{x,x+dx^{\mu}}\psi(x+dx^{\mu})=r_{x+dx^{\mu},x}V(x+dx^{\mu},x)\psi(x+dx^{\mu})_{x}.
              \end{equation}

              Use of this in  Lagrangians and the requirement  that terms in the Lagrangians are limited to those that are invariant under local $U(n)$ transformations, gives the result that QED and other Lagrangians contain an extra term of the form $g_{r}\bar{\psi}\gamma^{\mu}A_{\mu}\psi.$ Here $g_{r}$ is a coupling constant and $A_{\mu}(x)=\partial_{\mu,x}\Theta.$ A mass term for $\Theta$ may be present as it is not excluded by local $U(n)$ invariance.

               Abelian gauge theories are a good example of how this works.  Here the vector spaces are two dimensional and the expanded gauge group is $GL(1,C).$  The covariant derivative in the Dirac Lagrangian, \begin{equation}\label{DirL}L(\psi,\bar{\psi})=i\bar{\psi}\gamma^{\mu} D_{\mu}\psi-m\bar{\psi}\psi,\end{equation} is given by Eq. \ref{Dmupsi}.

               The requirement that all terms in the Lagrangian \cite{Montvay,Cheng} be  invariant under local $U(1)$ gauge transformations leads to the requirement that \cite{Cheng} \begin{equation}\label{DpLLD} D^{\prime}_{\mu,x}\Lambda(x)\psi=\Lambda(x)D_{\mu,x}\psi\end{equation} Here $\Lambda(x)=e^{i\phi(x)}$ is a local $U(1)$ gauge transformation.

               Setting \begin{equation}\label{rxdxmu}r_{x+dx^{\mu},x}=e^{g_{r}A_{\mu}(x)dx}\end{equation} and
               \begin{equation}\label{VgB}V(x+dx^{\mu},x)=e^{ig_{i}B_{\mu}(x)dx}\end{equation} in the covariant derivatives,  expanding the exponentials to first order and using  Eq. \ref{DpLLD} gives \begin{equation}\label{ABpAB}\begin{array}{c}A^{\prime}_{\mu}(x) =A_{\mu}(x)\\\\B^{\prime}_{\mu}(x)=B_{\mu}(x)-\frac{1}{g_{i}} \partial_{\mu,x}\phi(x).\end{array}
               \end{equation}Coupling constants $g_{r}$ and $g_{i}$ have been added for the $A$ and $B$ fields. $D^{\prime}_{\mu,x}$ is obtained from $D_{\mu,x}$ by replacing unprimed $A$ and $B$ fields with primed ones.

               Use of these results in the Dirac Lagrangian and adding a Yang Mills term for the $B$ field gives the result \begin{equation}\label{DirLAB}\begin{array}{l}L(\psi,\bar{\psi})= i\bar{\psi}\gamma^{\mu} (\partial_{\mu,x}+g_{r}A_{\mu}(x)+ig_{i}B_{\mu}(x))\psi-m\bar{\psi}\psi \\\\\hspace{.5cm}-\frac{1}{2}\lambda^{2}A^{\mu}(x)A_{\mu}(x)-\frac{1}{4}G_{I,\mu,\nu} G_{I}^{\mu,\nu}.\end{array}\end{equation}Here $B(x)$ is the usual photon field.

               This Lagrangian differs from the usual QED Lagrangian by the presence of an interaction term, $ig_{r}\bar{\psi}\gamma^{\mu}A_{\mu}(x)\psi(x),$ between the $A$ and matter fields and a mass term for the $A$ field. This result shows that the $\Theta$ field with $\vec{A}(x)=\nabla_{x}\Theta(x)$ is a boson field. The presence of a mass term for $\Theta$ indicates that the boson may have mass, however $\lambda=0$ is also possible.

               One property of $\Theta$ that one can be sure of is that the coupling constant, $g_{r},$ of $\vec{A}$ to matter fields must be very small compared to the fine structure constant. This is a consequence of the great accuracy of QED without the presence of the $\vec{A}$ field. It is also likely that $\Theta$ is a spin $0$ boson.  This is a consequence of the fact that the scaling factor, $r_{y,x},$ applies to  number structures and, by extension, vector spaces.

               At this point it is not known which physical field, if any, $\Theta$ represents. Candidates include the Higg's boson, gravity, dark matter, dark energy, etc. \cite{Li,Steinhardt}.  It is hoped in future work to determine if the boson field, $\Theta$ is any one of these fields or is something else.

              \section{The basic model}\label{BM}
              \subsection{General description}
              So far, the description of separate complex number structures, $\bar{C}_{x}$ and vector spaces at each point $x$ has been limited to  their use in  gauge theories. This suggests that one \emph{explore} some consequences of the expansion of this  to  a model of physics, geometry, and mathematics in which  the basic setup consists of  separate mathematical structures of different types associated with each  point, $x$, of a  space time manifold, $M$.

              The emphasis here is on exploration.  At present it is not known  if physics makes use of the $\Theta$ field. A first step in determining the relationship, if any, of $\Theta$ to physics is to explore the effects of $\Theta$ on physical and geometric entities.  It will be seen that $\Theta$ does effect theoretical descriptions of these entities.

              The types of mathematical structures assigned to each point, $x$, of $M$ include structures for numbers of different types (natural numbers, integers, rational, real, and complex) and any other systems that include numbers in their description. These include vector spaces, algebras of operators, group representations, etc. Each system type $S$ is included as a separate  structure, $\bar{S}_{x},$ at each $x$ that satisfies a set of  axioms \cite{Barwise,Keisler}  relevant to the structure type.

              In more detail a structure, $\bar{S}_{x},$ can be represented as\begin{equation}\label{Strx}
              \bar{S}_{x}=\{S_{x},Op_{x},Re_{x},K_{x}\}.\end{equation} $S_{x},Op_{x},Re_{x},K_{x}$ are the sets of base elements, the basic operations, the basic relations, and constants respectively. The scalars for $\bar{S}_{x}$ are those in $\bar{C}_{x},$ $\bar{R}_{x},$ or any other number type.

              The model of physics and mathematics considered here is different from that described by Tegmark \cite{Tegmark} in that physics systems are considered to be different from mathematical structures. It is not assumed that the universe of physical systems is mathematics as is done in \cite{Tegmark}. However, details on the differences between physical systems and mathematical structures must await future work.

               The association of mathematical structures at each point $x$ is combined with the idea that observers can be located at points of $M$.  This is clearly an idealization as observers are macroscopic objects that occupy finite regions of space time. This problem is easily remedied here by choosing $x$ to be any point in the region occupied by an observer. As far as scaling is concerned, it will turn out that the choice of the point $x$ in the observer occupied regions has no observable effect on scaling.

              The assumption is now made that the totality of mathematics directly available to the observer, $O_{x}$ at $x$, is limited to the mathematical system structures $\bar{S}_{x}$, at $x.$  This is based on the idea that all the mathematics that $O_{x}$ can use to make predictions, etc. is in his head. Mathematics  in a textbook that $O_{x}$ is reading or being communicated in a lecture is not available to $O_{x}$ until the information is recorded in $O_{x}'s$ brain. In other words, mathematics is local.

              The totality of mathematics available to $O_{x}$ is denoted by $\bigcup_{x}.$ As might be expected, $\bigcup_{x}$ contains both unscaled and scaled structures.  For each structure type $\bar{S}_{x}$ in $\bigcup_{x}.$ there are also scaled structures, $\bar{S}^{r_{y,x}}_{x},$ for each point $y.$ Examples are the scaled complex number structures, $\bar{C}^{r_{y,x}}_{x}$ Eq. \ref{Crx}, real number structures, $\bar{R}^{r_{y,x}}_{x},$ and vector spaces $\bar{V}^{r}_{x}$ Eq. \ref{Vrx}. Most of these scaled structures will not be discussed here as they are not needed for the purposes of this paper.

              \subsection{Comparison of Theory and Experiment}\label{CTE}
              One area where one would expect scaling to have an effect is in the comparison of theory predictions with one another or with  experimental results. Suppose an experiment carried out at $x$ gives a numerical result $a_{x}$ and the same experiment repeated at $y$ gives the numerical result\footnote{Here $x$ and $y$ are arbitrary locations in the finite space time regions occupied by the experiments or computations.} $b_{y}.$ With separate number structures at each point it follows that $a_{x}$ and $b_{y}$ are real number values in  $\bar{R}_{x}$ (or $\bar{C}_{x}$) and $\bar{R}_{y}$ (or $\bar{C}_{y}$) respectively.

              Comparison of these two results requires transfer of both results to one location so that they are number values within one structure and can be locally compared.  If scaling is absent, then the two numbers to be compared at site $x$ are $a_{x}$ and $b_{x}=F_{x,y}b_{y}.$ Since $b_{x}$ is the same number value in $\bar{C}_{x}$ as $b_{y}$ is in $\bar{C}_{y}$, the relationship between the experimental numerical results is the same for separate number structures at each point as it is for one common structure, $\bar{C}$ for all points. With separate number structures one compares $a_{x}$ to $b_{x}$.  With one number structure one compares $a_{x}$ with $b_{y}$ as both number values are in $\bar{C}.$ If one suppresses statistical and quantum mechanical uncertainties, one would expect to find $a_{x}=b_{x}$ for separate number structures and $a_{x}=b_{y}$ for one number structure.

              The situation is different if scaling is present. With separate structures at each point, one compares, at $x,$ $a_{x}$ with $r_{y,x}b_{x}$ where $r_{y,x}$ is the scaling factor from $\bar{C}_{x}$ to $\bar{C}_{y}.$   Since the same experiment is done at both $x$ an $y,$  one would expect these two numerical results to be equal. Again statistical and quantum mechanical uncertainties are suppressed. This is clearly not true as $a_{x}\neq r_{y,x}b_{x}$.   A similar result is obtained for a comparison at $y$ as $b_{y}\neq r_{x,y}a_{y}.$  Thus, with scaling, an observer at $x$ or $y$ or at any point would conclude that the results of these two experiments are not equal.

              This description of comparison of theory with experiment causes problems for number scaling. The reason is that physics gives no hint of such inconsistencies.

              However, there is, in fact, no problem because scaling plays no role in the comparison of experimental results or theory predictions with one another or with experiment.  The reason is that the above description of comparison is not correct.   No experiment or theory computation ever gives a number value directly as output.  Outputs of experiment or theory computations are instead physical systems in physical states that are \emph{interpreted} as numbers. If $\phi_{x}$ and $\psi_{y}$ are output states of computations or experiments at $x$ and $y$, then the numerical values in $\bar{R}_{x}$ and $\bar{R}_{y}$ associated with these results are given by interpretation maps $J_{x}$ and $J_{y}$ as $J_{x}\phi_{x}$ and $J_{y}\psi_{y}.$

\begin{figure}[ht]\begin{center}\rotatebox{270}
              {\resizebox{120pt}{120pt}{\includegraphics[40pt,300pt]
              [440pt,690pt]{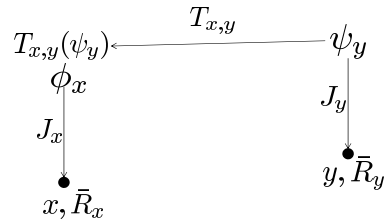}}}\end{center}\caption{Illustration of transmission and comparison of outputs of theory
              calculations or experiments.The theory or experiment output state, $\psi_{y}$ at $y$ corresponds to the number $J_{y}(\psi_{y})$ in $\bar{R}_{y}.$ Comparison of the output with the numerical result, $J_{x}(\phi_{x})$ of the theory or experiment output state, $\phi_{x}$ at $x$, requires transporting the information contained in $\psi_{y}$ to a common point, $x,$ for comparison. The numbers to be compared are $J_{x}(\phi_{x})$ and $J_{x}(T_{x,y}\psi_{y}).$}\label{ENS2}\end{figure}

              Comparison of these outputs requires  transmission of the information contained in these states  to a common point for comparison. Typically this is done by use of physical media such as light or sound. The transmission of the information must be such that information is not lost or distorted during transmission.  If $T_{x,y}(\psi_{y})$ denotes the state of the  physical system at $x$ that carries the information in  $\psi_{y}$ from $y$ to $x$, then comparison is between the number values $J_{x}T_{x,y}(\psi_{y})$ and $J_{x}\phi_{x}.$ There is no scaling involved in this comparison  of numerical results as the comparison  is done locally at some point and not at different points.   Figure \ref{ENS2} is a graphic illustration of this process of transmission and local comparison of experimental results.

              One might conclude from this that number scaling has no effect in physics and can be dispensed with. This is not the case.  Number scaling affects all theoretical descriptions of physical systems that involve derivatives or integrals over space or  time. For instance, suppose the theory prediction of the numerical outcome of the experiment at $y$ is described by an integral or derivative over space time or space or time.  Then, if scaling is present,  the predicted value to be compared with the experimental value, $J_{y}\psi_{y},$  is different from the predicted value if scaling is absent. Examples of this will be given in the following sections.

              \section{Effects of $\Theta$ on Physics}
              As was noted earlier a main goal is to determine the physical properties of  the field $\Theta$ and its relationship to other physical fields. So far one knows from the great accuracy of QED that the coupling of $\Theta$ to matter fields must be very small.  Also it is not known if $\Theta$ is massless or has a mass.

              Additional properties of the boson field, $\Theta,$ can be determined by examining the effect  of scaling on  physics. As seen in other work \cite{BenQT,BenSPIE2}, The requirement that mathematics is locally available means that one must address the question of how $O_{x}$ deals with mathematical descriptions of physical systems  that use integrals or derivatives over space time or space and/or time. The description of derivatives in gauge theories has already been described.  However, there are many other examples.

              A simple example of an integral over space is the description of wave packets in  nonrelativistic quantum mechanics as \begin{equation}\label{psiy}\psi=\int d^{3}y |y\rangle\langle y|\psi\rangle.\end{equation} In the usual description in quantum mechanics, all the vectors $|y\rangle\psi(y)$ belong to one Hilbert space. The addition of vectors implied in the integral has meaning as addition is defined within the Hilbert space.

               Eq. \ref{psiy} loses its meaning under the assumption of local availability of mathematics with separate Hilbert spaces and complex numbers, $\bar{H}_{y},\bar{C}_{y},$ at each point $y.$    The problem is that the definition of the Hilbert space containing $\psi$ is problematic. The reason is that there are separate complex number structures for each point, $y,$ instead of just one common structure.

               One approach is to start with structures, $\bar{H}_{y},\bar{C}_{y},$ for each $y$ where each $\bar{H}_{y}$ is one dimensional.  It contains the vectors, $c|y\rangle$ where $c$ is any complex number in $\bar{C}_{y}.$ One can form a direct sum Hilbert space provided the $\bar{C}_{y}$ are all mapped to a common structure.\footnote{The fact that direct sums of spaces over continuous variables cannot be defined is similar to the fact that space location states, $|x\rangle,$ are not proper eigenvectors of a Hilbert space. In keeping with usage, these problems are ignored here.}

              In the absence of scaling, the problem can be fixed by use of the unitary parallel transform operators, $U_{x,y},$ in Eq. \ref{UZW}  and $F_{x,y}$ in Eq. \ref{Fyx} to map $\bar{H}_{y}$  onto $\bar{H}_{y)x}$ and $\bar{C}_{y}$ to $\bar{C}_{x}.$ Then the spaces, $\bar{H}_{y)x},$  can be summed to create a single Hilbert space for wave packet vectors.  This is shown in  \begin{equation}\label{dirsum}\begin{array}{c} \bar{H}_{y}\\\bar{C}_{y}\end{array} \Rightarrow\begin{array}{c}\bar{H}_{y)x}\\\bar{C}_{x} \end{array}\Rightarrow\begin{array}{c}\bigoplus_{y}\bar{H}_{y)x} \\ \bar{C}_{x}\end{array}.\end{equation} The vectors in $\bar{H}_{y)x}$ all have the form $c|y\rangle$ where $c$ is the same complex number value in $\bar{C}_{x}$  as it is in $\bar{C}_{y}.$ A reference location, $x,$ is necessary in this setup.

               Wave packet states, as in Eq. \ref{psiy}, can be expressed in this Hilbert space as,\footnote{The action of $U_{y,x}$ is combined with that of $F_{y,x}$ in the mapping of vectors defined as the product of another vector and a scalar.}\begin{equation} \label{psixUF} \begin{array}{l}\psi_{x}=\int_{x}U_{x,y}(|y\rangle\psi(y)d^{3}y)=\int_{x}U_{x,y} (|y\rangle)F_{x,y}(\psi(y)d^{3}y)\\\\\hspace{1cm}=\int_{x}|y\rangle_{x}\psi(y)_{x}
              dy^{3}_{x}.\end{array}\end{equation}This is equivalent to the wave packet integral in the usual case with just one $\bar{C}$ and one $\bar{H}$ for all points of $M$. Also the probability of finding the system $\psi$ somewhere, given by \begin{equation}\label{norm}\langle\psi|\psi \rangle_{x}=\int_{x}F_{x,y}(|\psi(y)|^{2}d^{3}y)\end{equation} is the same as the value for just one $\bar{H}$ and $\bar{C}.$

              The same holds for the  expectation value, of the position operator, $\tilde{y},$ \begin{equation}\label{pos}\langle \psi |\tilde{y}|\psi\rangle_{x} =\int_{x}y_{x}|\psi(y)|^{2}_{x}d^{3}y_{x}.\end{equation}  This is the same real number in $\bar{C}_{x}$ as is the usual value in the case of just one $\bar{H}$ and one $\bar{C}.$

              The inclusion of scaling changes these results.   In this case, Eq. \ref{dirsum} is replaced by \begin{equation}\label{dirsumsc}\begin{array}{c}\bar{H}_{y}\\ \bar{C}_{y}\end{array}\Rightarrow \begin{array}{c}Z_{x,y}\bar{H}_{y} \\X_{x,y}\bar{C}_{y} \end{array}=\begin{array}{c} \bar{H}_{y)x}^{r}\\\bar{C}^{r}_{x}\end{array}\Rightarrow\begin{array}{c}\bigoplus_{y}
              \bar{H}^{r}_{y)x}\\\bar{C}_{x}\end{array}.\end{equation} Here $r=r_{y,x}$ is the scaling factor, $Z_{x,y}=W_{y,x}U_{x,y},$ Eq. \ref{UZW}  and $X_{x,y}=Y_{y,x}F_{x,y},$ Eq. \ref{FXY}.  $\bar{H}^{r}_{y)x}$  is given in footnote \ref{Hrx} and $\bar{C}^{r}_{x}$ is given by Eq. \ref{Crx}. Vectors in $\bar{H}^{r}_{y)x}$ (one dimensional) have the form $r_{y,x}c|y\rangle_{x}$ with $c$ the same number value in both $\bar{C}_{y}$ and $\bar{C}_{x}.$ The right hand implication expresses the observation that the different $\bar{C}^{r}_{x}$ can all be mapped with scaling onto $\bar{C}_{x}.$

              This result shows that  inclusion of scaling requires that the term $|y\rangle_{x}\psi(y)_{x}$  in Eq. \ref{psixUF} be replaced by $r_{y,x}|y\rangle_{x}\psi(y)_{x}.$ The resultant wave packet integral is\begin{equation}\label{psiscx}\psi^{\Theta}_{x}=\int_{x}r_{y,x} |y\rangle_{x}\psi(y)_{x}dy^{3}_{x}=\int_{x}e^{\Theta(y)_{x}-\Theta(x)} |y\rangle_{x}\psi(y)_{x}dy^{3}_{x}.\end{equation} Also the  probability and position expectation values become  \begin{equation}\label{normSc}\langle\psi|\psi\rangle^{\Theta}_{x}=\int_{x} e^{\Theta(y)_{x}-\Theta(x)} |\psi(y)|^{2}_{x}dy^{3}_{x}\end{equation}
              and \begin{equation}\label{psiposscx}\langle\psi|\tilde{y}|\psi\rangle^{\Theta}_{x} =\int_{x}e^{\Theta(y)_{x}-\Theta(x)} y_{x}|\psi(y)|^{2}_{x}dy^{3}_{x}.\end{equation}
              The subscripts, $x,$ indicate that all numerical values, states, and operations in these equations belong to $\bar{H}_{x}$ and $\bar{C}_{x}$ in $\bigcup_{x}.$

            The reference point $x$ for these integrals can be changed from $x$ to another point $z$ by applying an external scaling factor that reflects the change.  \begin{equation}\label{extintpsi} \begin{array}{l}\psi^{\Theta}_{z}= e^{\Theta(x)_{z}-\Theta(z)}U_{z,x} (\int_{x}e_{x}^{\Theta(y)_{x}-\Theta(x)} |y\rangle_{x}\psi(y)_{x}dy_{x})\\\\ \hspace{1cm}=\int_{z}e_{z}^{\Theta(x)_{z} -\Theta(z)+\Theta(y)_{z}-\Theta(x)_{z}} |y\rangle_{z}\psi(y)_{z}dy_{z}
            \\\\\hspace{2cm}=\int_{z}e_{z}^{\Theta(y)_{z} -\Theta(z)}|y\rangle_{z} \psi(y)_{z}dy_{z}\end{array}\end{equation} and for the expectation values, \begin{equation}\label{psinormzz}\begin{array}{l}\langle\psi|\psi\rangle^{\Theta}_{z}
            =e^{\Theta(x)_{z}-\Theta(z)}F_{z,x} (\int_{x}e_{x}^{\Theta(y)_{x}-\Theta(x)} |\psi(y)|^{2}_{x}dy_{x})\\\\\hspace{1cm}=\int_{z}e_{z}^{\Theta(y)_{z} -\Theta(z)}|\psi(y)|^{2}_{z}dy_{z}\end{array}\end{equation} and \begin{equation}\begin{array}{l}
            \langle\psi|\tilde{y}|\psi\rangle^{\Theta}_{z}=e^{\Theta(x)_{z}-\Theta(z)}F_{z,x} (\int_{x}e_{x}^{\Theta(y)_{x}-\Theta(x)} |\psi(y)|^{2}_{x}dy_{x})\\\\\hspace{1cm}=\int_{z}e_{z}^{\Theta(y)_{z} -\Theta(z)}y_{z}|\psi(y)|^{2}_{z}dy_{z}.\end{array}\end{equation} It is clear from these expressions that the dependence of these values on the reference point, $z,$ is given by the term $\Theta(z)$ in the exponent of the scaling factor.

              The usual expression for the momentum operator, \begin{equation}\label{pus}
             \tilde{p}=\hbar \tilde{k}=i\hbar \sum_{j}\partial_{j,y},\end{equation}is changed in the presence of separate number structures and   scaling. As was seen in gauge theories, the  usual expression for the  derivative,  \begin{equation}\label{derpsi} \partial_{y,j}\psi=\lim_{dy^{j}\rightarrow 0}\frac{\psi(y+dy^{j}) -\psi(y)}{dy^{j}},\end{equation}  makes no sense  because $\psi(y+dy^{j})$ is in $\bar{C}_{y+dy^{j}}$ and $\psi(y)$ is in $\bar{C}_{y}$.

             This can be remedied by parallel transporting $\psi(y+dy^{j})$ to $\bar{C}_{y}$.  The derivative becomes  \begin{equation}\label{parpri}\partial'_{y,j} \psi=\lim_{dy^{j}\rightarrow 0}\frac{\psi(y+dy^{j})_{y}-\psi(y)}{dy^{j}}. \end{equation}Here $\psi(y+dy^{j})_{y}=F_{y,y+dy^{j}}\psi(y+dy^{j})$ is the same number value in $\bar{C}_{y}$ as $\psi(y+dy^{j})$ is in $\bar{C}_{y+dy^{j}}.$  The prime on the derivative  refers to its definition within $\bar{C}_{y}$.    $F_{y,y+dy^{j}}$ is the parallel transport operator, Eq. \ref{Fyx}, for number structures.

             Inclusion of scaling follows the description of the covariant derivative in gauge theory \cite{Montvay,Cheng}. Here $\partial'_{y,j}\psi$ becomes $D_{y,j}\psi$ where\begin{equation}\label{derDyj}D_{y,j}\psi =\lim_{dy^{j}\rightarrow 0}\frac{r_{y+dy^{j},y}\psi(y+dy^{j})_{y} -\psi(y)}{dy^{j}}.\end{equation}  Using \begin{equation}\label{rdyy}r_{y+dy^{j},y}=e^{\Theta(y+dy^{j})-\Theta(y)}=
             e^{\partial_{y,j}\Theta dy^{j}}\end{equation}as the scaling factor, and expanding the exponential to first order gives \begin{equation} \label{defDyj}D_{y,j}=\partial^{\prime}_{y,j}+ \partial_{y,j}\Theta =\partial_{y,j}+A_{j}(y).\end{equation}  The prime on the derivative has been dropped because it has no effect on the value of the derivative.

             This shows that, in the presence of scaling, the momentum operator, \begin{equation}\label{momsc} \tilde{p}_{x}=i\hbar\sum_{j}D_{x,j},\end{equation} is similar to the expression for the canonical momentum for the electromagnetic field. However it seems that Eq. \ref{momsc} must also be used for the actual physical momentum of a quantum system. In this case the kinetic energy  component of a Hamiltonian for a system is \begin{equation} \label{KESc} \tilde{K}=\frac{-\hbar^{2}D^{2}_{x}}{2m}.\end{equation}

              Another area in which scaling affects physics is  the derivation of equations of motion from the action.  Scaling would be expected to have an effect since the action is an integral over space and time or space time of the Lagrangian density.

              A simple example consists of the derivation \cite{Baez} of the equation of motion of a classical system from the action\begin{equation}\label{clS}S(\gamma)=\int_{0}^{t}L(\gamma(s),\dot{\gamma}(s))ds.\end{equation} Here $\gamma$ is the path taken by the particle and $\gamma(s)$ is the particle position at time $s.$ With no scaling but separate mathematical systems at each point the integrand must be parallel transported to some common point, $x,$ for the integral to make sense. The result is \begin{equation}\label{clSx}S(\gamma)_{x}=\int_{x,0}^{t}L(\gamma(s),
              \dot{\gamma}(s))_{x}ds.\end{equation} With scaling included the action becomes, \begin{equation}\label{ScclS}S^{\Theta}(\gamma)_{x}=\int_{x,0}^{t}e^{\Theta(\gamma(s))
              -\Theta(x)}L(\gamma(s),\dot{\gamma}(s))_{x}ds.\end{equation} The exponential factor accounts for scaling in transferral of the integrand from $\gamma(s)$ to $x.$

              For the unscaled action the  Euler Lagrange equations give \cite{Baez} Newton's equation of motion as \begin{equation}\label{EL}
              \frac{dp(t)}{dt}\equiv\frac{d}{dt}(\frac{dL}{d\dot{\gamma}(t)})=\frac{dL} {d\gamma(t)}\equiv F.\end{equation} Here $p(t)$ and $F$ are the particle momentum and force on the particle and $L$ is the Lagrangian. With scaling included one obtains, \begin{equation}\label{ScEL}\frac{dp(t)}{dt}= (\frac{d}{dt}+\frac{d}{dt}\Theta(\gamma(t)))(\frac{dL}{d\dot{\gamma}(t)})=
              (\frac{d\Theta}{d\gamma})L+\frac{dL} {d\gamma(t)}=F.\end{equation} This result is obtained by letting $Z(\Theta,\gamma,\dot{\gamma})$ be the integrand of Eq. \ref{ScclS}, carrying out the derivatives in the Euler Lagrange equation, \begin{equation}\label{ELZ}
              \frac{d}{dt}(\frac{dZ}{d\dot{\gamma}})=\frac{dZ}{d\gamma},\end{equation} and cancelling out the common exponential factor. There is no variation with respect to $\Theta$. One can also set\begin{equation}\label{cmom}\frac{d}{dt}(\Theta(\gamma(t))= \sum_{j}\frac{d\Theta}{d\gamma^{j}}\dot{\gamma}^{j}(t) =\sum_{j}A_{j}(\gamma(t))\dot{\gamma}^{j}(t).\end{equation}

              The presence of $\Theta$ results in two new terms in the equation of motion.  The  new term $(\frac{d\Theta}{d\gamma})L$  contributes to the force on the particle. If one sets $\frac{d\Theta}{d\gamma}=\vec{A}(\gamma(t))$, then the extra force term becomes $\vec{A}(\gamma(t))L.$ This term is always present if there is scaling and $L\neq 0.$ This includes the case where $L=(1/2)m\dot{\gamma}^{2}$  is the usual kinetic energy Lagrangian for which the usual force term $dL/d\gamma=0.$

              The other new term contributes to the time rate change of the momentum.  It is always present if there is scaling and there are terms in the Lagrangian that include $\dot{\gamma}.$

              The field $\Theta$ also appears in many other equations of motion. As an example, for the the Dirac Lagrangian  density, \begin{equation}\label{DL}L(\bar{\psi}(y),\partial_{\mu}\psi(y))= \bar{\psi}(y)i\gamma^{\mu}\partial_{\mu,y}\psi(y)-m\bar{\psi}(y)\psi(y),\end{equation} the action, with number scaling included, is given by \begin{equation}\label{SScx}S^{\Theta}(\psi)_{x}= \int_{x} e^{\Theta(y)_{x}-\Theta(x)}L(\bar{\psi}(y),D_{\mu,y}\psi(y))d^{4}y.\end{equation} The subscript $x$ means that the integral is evaluated at $x$.

              Variation of the action with respect to $\psi$ or $\bar{\psi}$ gives equations of motion where the presence of $\Theta$ shows up only in the derivative, $D_{\mu,y}.$  The exponential factor multiplying the Lagrangian has no effect.  This is a consequence of the fact that it does not depend on $\psi.$  The result gives the Dirac Lagrangian with $D_{\mu,y}$ replacing $\partial_{\mu,y}$ as in \begin{equation}\label{DLD}i\gamma^{\mu}D_{\mu,y}\psi(y)-m\psi(y)=0.
              \end{equation}  Here \begin{equation}\label{Dmuy}D_{\mu,y}\psi(y)=(\partial_{\mu,y}+gA_{\mu}(y)) \psi(y)\end{equation} and $g$ is a coupling constant.

              A common feature of effects of scaling is that scaling depends only  on the difference between values of $\Theta$ at different points. The effects are invariant under changing the value of $\Theta(x)$ everywhere by a constant.  This follows from $\Theta(y)_{x}+c-(\Theta(x)+c)=\Theta(y)_{x}-\Theta(x).$

              There are many other examples of the effect of scaling in quantum physics that could be given.  However these are sufficient to show that the local availability of mathematics, without scaling, affects the form but not the values of theoretical predictions and descriptions. Predictions under the local availability of mathematics with no scaling are the same  as with one mathematical system of each type for all points of $M$.

              This is not the case if scaling is included. Predictions of values of physical quantities with scaling are different from the values without scaling.

              \section{Restrictions on $\Theta$}\label{RT}

              These few examples, and many more that can be constructed, show that the presence of scaling caused by the boson field $\Theta$  does affect theoretical predictions of properties of systems. Simple examples include the probability, at $x$  of finding the system $\psi$ somewhere,  given by Eq. \ref{normSc} as
              \begin{equation}\label{ProbT}\langle\psi|\psi\rangle^{\Theta}_{x}=\int_{x}e^{\Theta(y)_{x}- \Theta(x)}|\psi(y)|^{2}_{x} d^{3}y_{x},\end{equation}and the position expectation value, given by Eq. \ref{psiposscx} as \begin{equation}\label{PosT} \langle\psi|\tilde{y}|\psi\rangle^{\Theta}_{x}=\int_{x} e^{\Theta(y)_{x}-\Theta(x)}y|\psi(y)|^{2}_{x}d^{3}y_{x}\end{equation}

              All experimental tests of these properties of quantum systems, done so far, show no effect of the presence of  $\Theta.$ The probability of finding a system in state $\psi$ somewhere is equal to one and is independent of where the probability is calculated. Similarly there is no experimental evidence for the presence of $\Theta$ in comparing predicted expectation values with those from experiment. It follows that \begin{equation}\label{Prob}
              \int_{x}e^{\Theta(y)_{x}- \Theta(x)}|\psi(y)|^{2}_{x} d^{3}y_{x}- \int_{x}|\psi(y)|^{2}_{x} d^{3}y_{x}\simeq 0\end{equation}and\begin{equation}\label{Pos}
              \int_{x}e^{\Theta(y)_{x}- \Theta(x)}y|\psi(y)|^{2}_{x} d^{3}y_{x}- \int_{x}y|\psi(y)|^{2}_{x} d^{3}y_{x}\simeq 0.\end{equation}Here $\simeq 0$ means, equals $0$ to within experimental error.

              An important aspect of these and other experimental comparisons of theory with experiment in which systems are prepared in some state and their properties measured, is that the systems occupy  relatively small regions of space and time. It follows that the integrals over all space in the expectation values of Eqs. \ref{Prob} and \ref{Pos} can be replaced by integrations over finite volumes, $W.$ $W$ is determined by the requirement that the values of the integrals over all points outside $W$ are too small to be detected experimentally.

              This replacement of infinite space or space time integration volumes by finite ones holds for all theory experiment comparisons in which states are prepared by  observers at some location, $x$ and measured at some  location $y$ and the theory computations done at location $z$. The preparation, measurement and computation volumes of space time are all finite. This is the case even for statistical comparisons of theory with experiment.  If theory experiment comparison requires comparison of theory with the average value obtained from $n$ repetitions of an experiment, then the total space time volume required is roughly $n$ times that for a single experiment.

              Let $Z$ be a region of space time that includes all space and time points that are accessible to observers for preparing systems in states and carrying out measurements on the prepared systems, or for making computations. The fact that the effect of $\Theta$ is not observed, means that for all pairs, $y,x$ of points in $Z$,\footnote{This requirement is different from the fact that  scaling plays no role in the comparison of theory with experiment as discussed in subsection \ref{CTE}. Here one is referring to the inclusion of  scaling factors inside integrals over space or space and time. These internal scaling factors are present in all theoretical predictions that involve space  and/or time integrals.}  \begin{equation}\label{ReqT}\Theta(y)_{x}-\Theta(x)\simeq 0.\end{equation} Here  $x$ denotes the location of any actual or potential observer location and $y$ is in the sensitive volume of any actual or potential experiment or of any computation. The condition, $\simeq 0,$ means that the effect of $\Theta,$ if any, is too small to be observed experimentally.

              The anthropic principle \cite{Anth} plays a role here. It states that the physical laws and properties of physical systems must be such that there exists a region $Z$ of space and time in which, we, as intelligent observers, can exist, make theoretical predictions, and carry out experiments to test the predictions.

              On a local scale, the region $Z$ is large. It includes the earth as all experiments and calculations carried out so far have been by observers on or very near the earth.  However $Z$ must also include locations for which there is a potential for observers to exist, carry out experiments, and communicate with terrestrial observers. It follows that $Z$ must include the solar system  as the potential for observers to carry out experiments on solar system planets and in orbit around planets must be included.

              A generous estimate of the spatial extent of  $Z$ is as a sphere of a radius of a few light years centered on the solar system. The reason it is not larger is that the time required to establish multiple round trip communications with intelligent beings in regions outside $Z$, if any exist, and  to discover that there is no effect of $\Theta$ at these distant locations, is prohibitively long.

              This rough estimate of the size of $Z$ gets some support from another  estimate of the size of the larger region in which it is possible for us, as terrestrial observers, to just determine if intelligent beings  even exist. This region is estimated \cite{AmerSci}  to be a region that includes stars that are at most, about 1,200 light years distant.  If intelligent beings exist anywhere outside this region, we will never become aware of their existence.

              It is difficult to set a restriction on the time range of $Z$. It certainly includes the present and recent  past. How far it extends into the future is unknown.  For these reasons no time range will be assigned to $Z$. The only restriction is that it includes the present and recent past.

               The important point here is that the restrictions on values of $\Theta,$ in Eq. \ref{ReqT}, are limited to points in $Z$ and that $Z$ is small on a cosmological scale.  The exact size of $Z$ is not relevant. So far there are no restrictions on the values of $\Theta$ at locations outside $Z$.  This includes the effect of scaling on theoretical descriptions of very large systems or systems at cosmological distances.

              \section{Effects of $\Theta$  on Geometry}\label{ETG}
              As noted, the result that $\Theta(y)_{x}-\Theta(x)\simeq 0$ locally  for all points in $Z$ does not exclude the possibility that the boson field affects properties of physical and geometrical structures that are large on a cosmological scale. The restrictions also do not apply if the location, $y$, of some event is far away from us, as observers of the event from locations in $Z$.  It is also possible that the restriction does not apply for $x$ and $y$ in regions of size $Z$ that are cosmologically far away from us.

             For these reasons it is worthwhile to investigate the effects of $\Theta$ on  geometric quantities, particularly over large distances. The geometry to be investigated is that  of the basic model, Section \ref{BM}, which is  the assignment of separate number structures, $\bar{C}_{x}$ and $\bar{R}_{x}$, at each point, $x,$ of a space time manifold, $M$.  As noted, scaling affects all quantities that involve integrals or derivatives over space time or space and/or time. It also is used in the change of reference points for observers providing a  mathematical description of physical and geometric properties.

             \subsection{Effect of Scaling on the Line Element, $ds^{2}_{x}$}
            It is useful to begin with a description of the effect of scaling on the line element, $ds^{2}_{x}=\sum_{\mu,\nu}g_{\mu,\nu}(x)dx^{\mu}dx^{\nu}.$
            The subscript $x$ means that this is the line element at point $x$ of $M$. Under the usual setup, the metric tensor, $g_{\mu,\nu}(x),$ takes values in $\bar{R}$ for all values of $x,$ and $dx^{2}$ is based on $\bar{R}$ for all values of $x.$ This follows from the association of just one real number structure  with all points of $M$. The components, $dx^{\mu},$ of the vector are elements of an $n$ tuple of numbers in $\bar{R}^{n}.$ ($M$ is assumed to be $n$ dimensional.) This represents a coordinate system that is  valid locally at $x.$

            Here the setup is different in that separate real number structures $\bar{R}_{x}$ are associated with each point $x$ of $M$. In this case the metric tensor components, $g_{\mu,\nu}(x),$ are values in $\bar{R}_{x}.$ $ds^{2}_{x}$ is based on $\bar{R}_{x}$ and the components, $dx^{\mu},$ are elements of an n-tuple in $\bar{R}^{n}_{x}.$

            This description is satisfactory for an observer at $x$. However an observer, $O_{z},$ at another reference point, $z,$ uses mathematics based on $\bar{R}_{z}.$  For $O_{z},$ the description of $ds^{2}_{x}$ must be based on $\bar{R}_{z},$ not $\bar{R}_{x}.$

            This is done by mapping values of $ds^{2}_{x}$  into $\bar{R}_{z}.$  In the absence of scaling, parallel transformations are sufficient for the mapping of  $ds^{2}_{x}$ to $z$.  One obtains
            \begin{equation}\label{dx2nSc}(ds^{2}_{x})_{z}=F_{z,x}ds^{2}_{x}=F_{z,x}(g_{\mu,\nu}(x)) F_{z,x}(dx^{\mu}dx^{\nu})=g_{\mu,\nu}(x)_{z}dx^{\mu}_{z}dx^{\nu}_{z}.\end{equation} Here $g_{\mu,\nu}(x)_{z},$ $dx^{\mu}_{z},$ and $dx^{\nu}_{z}$ denote the same values in $\bar{R}_{z}$ as $g_{\mu,\nu}(x),$ $dx^{\mu},$ and $dx^{\nu}$ are\footnote{It is good to emphasize that parallel  transformations of number values from one number structure to another are not the same as space or time  translations of events or physical objects from one point to another.} in $\bar{R}_{x}.$

            It follows that $(ds^{2}_{x})_{z}$ has the same value in $\bar{R}_{z}$ as $ds^{2}_{x}$ does in $\bar{R}_{x}.$ This shows that, in the absence of scaling,  the value of the line element $ds^{2}_{x}$ at point $x$ is independent of which number structure is used for the value of $ds^{2}_{x}$. Replacement of a single $\bar{R}$ by separate $\bar{R}_{x}$ at each $x$ has no effect.

            This is no longer the case if scaling is included. Then the value of $ds^{2}_{x}$ at $z$ is multiplied by the scaling factor from $x$ to $z.$ The result is \begin{equation}\label{dx2Sc}\begin{array}{l}(ds^{2}_{x})^{\Theta}_{z}=r_{x,z}F_{z,x}ds^{2}_{x} =e^{\Theta(x)_{z}-\Theta(z)}(ds^{2}_{x})_{z}\\\\\hspace{1cm}=e^{\Theta(x)_{z}- \Theta(z)}g_{\mu,\nu}(x)_{z}dx^{\mu}_{z}dx^{\nu}_{z}.\end{array}\end{equation}
             As before, $F_{z,x}$ parallel transforms  numerical quantities from $x$ to $z,$ and $e^{\Theta(x)_{z} -\Theta(z)}$ gives the effect of scaling on the reference point change.

             Eq. \ref{dx2Sc} shows that, relative to an observer at $z$, $(ds^{2}_{x})^{\Theta}_{z}$ depends on $x.$ For reasons discussed in subsection \ref{RT}, the scaling factor is  independent of all reference locations,  $z$, in $Z$.

             The expression,  $\exp{\Theta(x)_{z}-\Theta(z)},$ for the scaling factor, $r_{y,x},$ is valid if the vector field, $\vec{A}(x),$ appearing in the original definition of $r_{y,x},$ Eq. \ref{ryx}, is integrable. If $\vec{A}(x)$ is not integrable, then Eq. \ref{ryxAd} is used for the scaling factor and Eq. \ref{dx2Sc} for the scaled line element is replaced by \begin{equation}\label{lelemScA} (ds^{2}_{x})^{\Theta}_{z}= e^{\int_{z}\vec{A}(\gamma(s))\cdot\nabla_{s}\gamma ds} (ds^{2}_{x})_{z}.\end{equation} Here $\gamma$ is a path from $z$ to $x.$

             This introduces a complication in that the scaling factor depends on the path from $z$ to $x.$ For $\vec{A}$ not integrable, it is an open question whether this path dependence should remain or  be removed by carrying out  some type of path integration.

             The scaling of the line element $ds^{2}_{x}$ shows that, if $\vec{A}$ is integrable, then  the scaling factor is independent of the geometry in that it is independent of  the metric tensor.  It is the same for Riemann geometries as it is for Euclidean and Minkowski geometries. However, scaling influences geometries because many geometric properties have numerical values.

             Euclidean spaces present a simple example of the effect of scaling.  For these spaces the  line element at $x$, referenced with scaling to $z,$ is
             \begin{equation}\label{Edx2Scz}(ds^{2}_{x})^{\Theta}_{z}=r_{x,z}F_{z,x}\vec{dx}\cdot_{x}\vec{dx}=r_{x,z} \vec{dx}_{z}\cdot_{z} \vec{dx}_{z}=r_{x,z}\sum_{j=1}^{n}(dx^{j})^{2}_{z}.\end{equation} Here $\cdot_{x}$ and $\cdot_{z}$ denote $x$ and $z$ based scalar products of vectors on Euclidean $M$.

             This result can be described by replacement of the usual Euclidean metric tensor,
            \begin{equation}\label{gijeucl}g_{i,j}(x)=\delta_{i,j}\end{equation}  by an $x$ dependent tensor \begin{equation}\label{gijyx}g_{i,j}(x,z)=r_{x,z}(\delta_{i,j})_{z}=e^{\Theta(x)_{z}- \Theta(z)}(\delta_{i,j})_{z}. \end{equation} The tensor is still diagonal in the component indices. However for each reference location, $z$, it is a function of $x$. In Eq. \ref{gijeucl} the   delta function components are number values in $\bar{R}_{x}$. In Eq. \ref{Edx2Scz} they are values in $\bar{R}_{z}.$ The equations also show that $(\delta_{i,j})_{z}=F_{z,x}(\delta_{i,j})_{x}.$

            One result of scaling is that, in scaled Euclidean space,  coordinate systems are valid only locally.   To see this, let $CS_{x}$ be a coordinate system with origin at $x$. If $CS_{x}$ is valid globally, then coordinates of  points in $M$ are described by n-tuples of number values in $\bar{R}^{n}_{x}.$   As a geometric entity at $x,$  the line element, written out so that the scalar product operation in $CS_{x}$  and arithmetic operations in $\bar{R}_{x}$ are made explicit, is\begin{equation}\label{ds2expl}
            ds^{2}_{x}=\vec{dx}\cdot_{x}\vec{dx}=\sum_{i,j}(\delta_{i,j})_{x}\times_{x}dx^{i}
            \times_{x}dx^{j}.\end{equation}

            Let $CS_{z}$ be a  coordinate system  with origin at $z$. If $CS_{z}$ is valid globally, then coordinates of points in $M$  would correspond to n-tuples in $\bar{R}_{z}^{n}$ and the line element at $x$ would be described by Eq. \ref{ds2expl} with all subscripts $x$ replaced by $z.$ This contradicts Eq. \ref{Edx2Scz} which shows that the representation of $ds^{2}_{x}$ in $CS_{z}$ includes $r_{x,z}$ as a scaling factor. This shows that number tuples in $\bar{R}^{n}_{z}$ cannot be used to describe geometric elements at locations $x$ that are different from $z.$

            Another way to understand this is to note that the $n$ tuple, $\bar{R}^{n}_{x},$  used  in $CS_{x},$ corresponds to the $n$ tuple of scaled real number structures, $(\bar{R}^{r_{x,z}}_{z})^{n},$ at $z$.  Here $\bar{R}^{r_{x,z}}_{z}$, given by Eqs. \ref{Crx} and \ref{Crx1}  with $R$ replacing $C,$ is the scaled representation of $\bar{R}_{x}$ on $\bar{R}_{z}.$

            This would lead to a strange coordinate system where the coordinates of each point, relative to the origin include a location dependent scaling factor. For a coordinate system with origin at $z,$ the coordinates of each point $x$ would be an $n$-tuple of numbers in $(\bar{R}^{r_{x,z}}_{z})^{n}.$ Note the $x$ dependence of the scaling factor, $r_{x,z}.$

            Fortunately, this representation of $CS_{x}$ on $CS_{z}$ can be replaced  by $r_{x,z}\bar{R}^{n}_{z}$ provided the scaling of operations in $(\bar{R}^{r_{x,z}}_{z})^{n}$ is accounted for.  As a specific example, the scaled representation, at $z$, of $ds^{2}_{x},$  as shown in Eq. \ref{ds2expl}, is given by \begin{equation}\label{ds2Texpl}\begin{array}{l}(ds^{2}_{x})^{\Theta}_{z}=\vec{dx}^{r}_{z} (\cdot)^{r}_{z}\vec{dx}^{r}_{z}=\sum_{i,j}(\delta_{i,j})^{r}_{z}\times^{r}_{z} (dx^{i})^{r}_{z}\times^{r}_{z}(dx^{j})^{r}_{z}\\\\\hspace{1cm}=\sum_{i,j}r\times_{z}
            (\delta_{i,j})_{z}\frac{\times_{z}}{r}(r\times_{z}dx^{j}_{z})\frac{\times_{z}}{r}(r\times_{z}
            dx^{j}_{z})\\\\\hspace{1.5cm}=r\sum_{i,j}(\delta_{i,j})_{z}dx^{i}_{z}dx^{j}_{z} =r(ds^{2}_{x})_{z}.\end{array}\end{equation} To save on notation, $r_{x,z}$ is denoted by $r.$ The superscript $r$ and subscript $z$ denote membership in $\bar{R}^{r_{x,z}}_{z}.$

             This result is the same as that in Eq. \ref{Edx2Scz} for the scaled line element. It shows in detail how one can change the reference point of a geometric element from $x$ to $z$ by changing the reference points, with scaling, of the components of the geometric element. However one must also change the reference points of the multiplication operations.  As Eq. \ref{Crx} shows, scaling of the operations must be included with these changes.

             This result shows that reference point change, with scaling, from $x$ to $z$, of the components of a geometric entity and combining the components after change, gives the same result as first combining the components, and then changing the reference point of the resultant entity. However, this is the case only if the operations used in combining the components are scaled.

            The description for  Euclidean space can be extended to  $3+1$ space time which is the venue for special relativity. For the coordinate representation, $t,\vec{x},$  the metric tensor at $x$ in the $+,-,-,-$ representation  is
            \begin{equation}\label{etamunu}\eta_{\mu,\nu}(x)_{x}=1_{x}-(\delta_{i,j})_{x}.\end{equation}This can be used to write the line element  at $x=r,\theta,\phi,t$ as\begin{equation}\label{ds2SR}ds^{2}_{x} =\eta_{\mu,\nu}(x)_{x}dx^{\mu}dx^{\nu}=c^{2}_{x}dt^{2}_{x}-\vec{dx}^{2}= c^{2}_{x}dt^{2}_{x}-(dr^{2}_{x}+r^{2}_{x}d\Omega^{2}).\end{equation}

            With separate real number structures, $\bar{R}_{x},$ at each point, $x$, and with scaling included, the line element at $x,$ referenced to $z,$  is given by Eq. \ref{dx2Sc} with $g_{\mu,\nu}(x)$ replaced by $\eta_{\mu,\nu}(x).$ The result is \begin{equation}\label{dx2ScSR}\begin{array}{l}(ds^{2}_{x})^{\Theta}_{z} =\eta^{\Theta}_{\mu,\nu}(x,z)dx^{\mu}_{z}dx^{\nu}_{z}=e^{\Theta(x)_{z}- \Theta(z)}\eta_{\mu,\nu}(x)_{z}dx^{\mu}_{z}dx^{\nu}_{z}\\\\\hspace{1cm}=e^{\Theta(x)_{z}- \Theta(z)}(c^{2}dt^{2}-\vec{dx}^{2})_{z}.\end{array}\end{equation}Here $\eta^{\Theta}_{\mu,\nu}(x,z) =e^{\Theta(x)_{z}- \Theta(z)}\eta_{\mu,\nu}(x)_{z}$ can be regarded as the scaled metric tensor for special relativity. $(c^{2}dt^{2}-\vec{dx}^{2})_{z}$ denotes the line element at $x,$ parallel transformed to $z.$

            Eq. \ref{dx2ScSR} shows that the same scaling factor multiplies both the time and space line elements. It follows that the  sign of $ds^{2}$ is unaffected by scaling.  If $ds^{2}_{x}>0$, (time like), then $(ds^{2}_{x})^{\Theta}_{z}>0.$ If $ds^{2}_{x}=0$, (light like), so is $(ds^{2}_{x})^{\Theta}_{z}.$ If $ds^{2}_{x}<0$, (space like), so is $(ds^{2}_{x})^{\Theta}_{z}.$

            These results also show that the geodesic  distance between two points, determined by photons, is $0$ and that the value is independent of scaling.  The boson field, $\Theta$ has no effect on this distance.  This is not the case if $ds^{2}_{x}\neq 0.$ Then the value of $(ds^{2})^{\Theta}_{z}$ does depend on  $\Theta.$

            This is a consequence of the fact that $0$ is the only number value unaffected by scaling.  In this sense, it can be regarded as a "number vacuum" as it is invariant  under scaling.

            For an observer at a space time location $z=t,\vec{z},$ all theory expressions are based on $\bar{R}_{z}$ and the mathematics at $z.$ Eq. \ref{dx2ScSR} is an example of this for the scaled line element.  These expressions all refer to a single reference time point $t.$  However observers, along with all other physical systems, move on a world line in space time.  It follows that the reference point $z$ is changing with time.

            This can be accounted for by letting $p(\tau)$ be the world line of an observer where $\tau$ is the proper time for the observer and $p(\tau)$ is the observer's location at time $\tau.$ The line element at $x,$ referenced to the location of an observer at time $\tau,$ is defined by replacement of $z$ by $p(\tau)$ to obtain \begin{equation}\label{dx2ScSRp}(ds^{2}_{x})^{\Theta}_{p(\tau)} =\eta^{\Theta}_{\mu,\nu}(x,p(\tau))dx^{\mu}_{p(\tau)}dx^{\nu}_{p(\tau)}=e^{\Theta(x)_{p(\tau)}- \Theta(p(\tau))}(ds^{2}_{x})_{p(\tau)}.\end{equation} $(ds^{2}_{x})_{p(\tau)}= F_{p(\tau),x}ds^{2}_{x}$ is the same number value in $\bar{R}_{p(\tau)}$ as $ds^{2}_{x}$ is in $\bar{R}_{x}.$

\begin{figure}[ht]\begin{center}\rotatebox{-90}
              {\resizebox{180pt}{180pt}{\includegraphics[-20pt,170pt]
              [440pt,590pt]{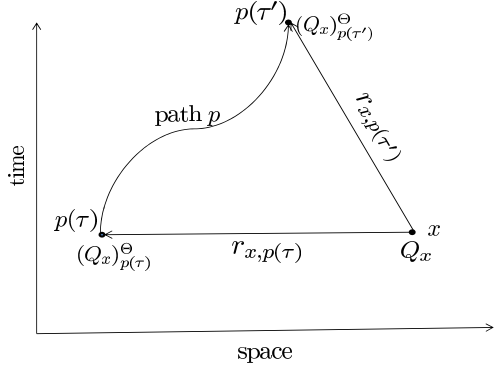}}}\end{center}\caption{Effect of change of reference point of quantity $Q_{x}$ at $x$ to two points $p(\tau)$ and $p(\tau')$ along an observers world line. $Q$ is used instead of the line element $ds^{2}$ to show that the scaling factors depends only on the location of the points, $x,p(\tau),$ and $p(\tau')$ in $M$. They are the same for all quantities $Q$ and for pure numbers at these locations.}\label{ENS3}\end{figure}

            Figure \ref{ENS3} shows the effect of world line motion on scaling for two times, $\tau$ and $\tau'.$ The quantity $Q$ is used in the figure to show that scaling depends only on point locations and not on the quantity being scaled.  It is the same for the line element as for any other quantity. 
            If the observer is in region $Z,$ as is assumed here, then the observer's world line is within  the accessible region $Z.$ In this case the motion of $O_{p(\tau)}$ along $p$ has no effect on the scaling as $\Theta(p(\tau'))\simeq \Theta(p(\tau))$ for any proper times $\tau, \tau'$ accessible to the observer.

             \subsection{Curve Lengths}\label{CL}
             The description of line elements on $M$ can be expanded to describe the effect of scaling on the length of curves on $M.$  Let $\gamma:[0,1]\rightarrow M$ be a smooth curve on $M$ parameterized by $s$ where $\gamma(0)=x$ and $\gamma(1)=y.$ The infinitesimal length of $\gamma$ at $s$ is represented by \begin{equation}\label{linelg}|\nabla_{s}\gamma|ds=(\nabla_{s} \gamma\cdot\nabla_{s}\gamma)^{1/2}ds=(\nabla_{s}\gamma ds\cdot\nabla_{s}\gamma ds)^{1/2}.\end{equation}Here  $\nabla_{s}\gamma$ is the gradient of $\gamma$ at $\gamma(s).$  $|\nabla_{s}\gamma |ds$ is the square root of the line element $dx^{2}=\vec{dx}\cdot\vec{dx}$ where $\vec{dx}$ is $\nabla_{s}\gamma ds.$

              Tangent spaces \cite{TangSp} are suitable venues for the description of line elements and curve lengths  for different geometries on $M.$  As an element of the tangent bundle on $M$,  the tangent space $T_{x}M,$ is the vector space of all vectors that are parallel to $M$ at $x.$ Coordinate systems on the tangent spaces can be used to give basis expansions of vectors in the tangent spaces.  For instance, coordinate system representations of $\vec{dx}$ and of $\nabla_{s}\gamma ds$ are given in $T_{x}M$ and $T_{\gamma(s)}M.$ The origins of these coordinate systems are at $x$ and $\gamma(s)$ respectively.

             The length of $\gamma$ is given by the line integral of $|\nabla_{s}\gamma|ds$ along $\gamma.$ In the usual setup with just one $\bar{R}$ for all points of $M$ the length is given by \begin{equation}\label{Lgamma} L(\gamma)=\int_{0}^{1}(\nabla_{s} \gamma\cdot\nabla_{s}\gamma)^{1/2}ds\end{equation} As such it is a vector in $T_{\gamma(s)}M.$

             This equation is valid everywhere in $M$. The reason is that  the real number structure, $\bar{R},$ and the mathematics based on them are the same everywhere. Thus an observer $O_{x}$ at any point $x$ with mathematics based on $\bar{R},$ accepts Eq. \ref{Lgamma} as valid. In particular the integrand, $(\nabla_{s} \gamma\cdot\nabla_{s}\gamma)^{1/2}$ is a number value in $\bar{R}$ for each value of $s$. The integral, as the limit of a sum of these values, makes sense as the number values to be added are all in $\bar{R}.$ The coordinate system representation of $\nabla_{s}\gamma ds$ as a vector in $T_{\gamma(s)}M,$ is based on $\bar{R}$ with vectors in this space corresponding to n-tuples of real numbers in $\bar{R}^{n}.$

             The setup changes for the basic model assumed here with assignment of separate real number structures to each point of $M.$ In this case, for each $s,$   the scalars for $T_{\gamma(s)}M$ are number values in $\bar{R}_{\gamma(s)}$ and  the coordinate system components of $\nabla_{s}\gamma$ are number tuples in $\bar{R}^{n}_{\gamma(s)}.$ Also $|\nabla_{s}\gamma|= (\nabla_{s}\gamma\cdot\nabla_{s}\gamma)^{1/2},$ is a number value in $\bar{R}_{\gamma(s)}.$ As a result the integral in Eq. \ref{Lgamma} does not make sense as it is the limit of a sum of real numbers, in different real number structures. Addition is defined only within real number structures, not between them.

             This can be fixed by choice of a reference point in $M$ and parallel transferring the integrand to  the number structure at the chosen point. For example if $x$ is the reference point, then the integrand values for each $s$ and the path integral  take values in $\bar{R}_{x}.$

             In the absence of scaling, one obtains \begin{equation} \label{Lgammax}L(\gamma)_{x}= \int_{x,0}^{1}F_{x,\gamma(s)}(|\nabla_{s} \gamma|ds)= \int_{x,0}^{1}|\nabla_{s} \gamma|_{x}ds_{x}.\end{equation}$F_{x,\gamma(s)}$  is the parallel transform operator, Eq. \ref{Fyx}, for number values from $\gamma(s)$ to $x,$ and $F_{x,\gamma(s)}|\nabla_{s} \gamma|=|\nabla_{s} \gamma|_{x}$ is the same number value in $\bar{R}_{x}$ as $|\nabla_{s} \gamma|$ is in $\bar{R}_{\gamma(s)}.$ Also $L(\gamma)_{x}$ is the same value in $\bar{R}_{x}$ as $L(\gamma)$ is in $\bar{R}$ in the usual setup with $\bar{R}$ common to all points of $M$.

             Scaling changes the value of $L(\gamma)_{x}.$ The reason is that transferral of integrand values of a common reference point includes scaling factors.  This is seen by noting that the length of $\gamma$ with scaling is given by \begin{equation} \label{LgScx}L(\gamma)^{\vec{A}}_{x}= \int_{x,0}^{1}r^{\gamma}_{\gamma(s),x}|\nabla_{s} \gamma|_{x}ds_{x}.\end{equation}In general $r^{\gamma}_{\gamma(s),x}$ is given by Eq. \ref{ryxAd} as \begin{equation}\label{rpathsx} r^{\gamma}_{\gamma(s),x} =\exp_{x}(\int_{x,0}^{s}\vec{A}(\gamma(u)) \cdot\nabla_{u}\gamma du).\end{equation}  The superscript $\gamma$ shows  the path dependence of the scaling factor .The subscript $x$ shows that the exponential and integral in the exponent are defined on $\bar{R}_{x}$ at $x.$ If $\vec{A}$ is the gradient of a scalar field, $\Theta,$  then the path dependence of the scaling factor disappears and  Eq. \ref{LgScx} becomes \begin{equation} \label{LgThx}L(\gamma)^{\Theta}_{x}= \int_{x,0}^{1}r_{\gamma(s),x}|\nabla_{s} \gamma|_{x}ds_{x}=\int_{x,0}^{1}e^{\Theta(\gamma(s))-\Theta(x)}|\nabla_{s} \gamma|_{x}ds_{x}.\end{equation}

            Note that in Eq. \ref{LgThx}  the scaling factor is inside the integral because it  depends on the value of the integration variable.  This type of scaling is called  \emph{internal scaling}. The reason is that it occurs inside a mathematical operation, such as integration or derivation over space, time, or  space time. It describes the transfer of the mathematical elements to a reference  point where the mathematical operation can combine the elements.

             This is distinguished from \emph{external scaling} which refers to the change of the reference point of a mathematical operation. This occurs outside the mathematical operation being considered. For example, changing the reference point of the scaled length, $L(\gamma)^{\Theta}_{x}$, of $\gamma$ from $x$ to the endpoint, $y,$ of $\gamma,$ is obtained by parallel transforming the integral of Eq. \ref{LgThx} to $y$ and multiplying by the scaling factor $r_{x,y}.$  One obtains \begin{equation} \label{LgThz}L(\gamma)^{\Theta}_{y}=r_{x,y}F_{y,x}L^{\Theta}(\gamma)_{x}=r_{x,y} (L^{\Theta}(\gamma)_{x})_{y} =e^{\Theta(x)_{y}-\Theta(y)} (L^{\Theta}(\gamma)_{x})_{y}. \end{equation} Here $(L^{\Theta}(\gamma)_{x})_{y}$ is the same number value in $\bar{R}_{y}$ as $L^{\Theta}(\gamma)_{x}$ is in $\bar{R}_{x}.$

             Writing $(L^{\Theta}(\gamma)_{x})_{y}$ as an integral and combining the scale factors gives
              \begin{equation}\label{LgThy1}\begin{array}{l}L^{\Theta}(\gamma)_{y}=r_{x,y}\int_{y,0}^{1} (r_{\gamma(s),x})_{y} |\nabla_{s}\gamma|_{y}ds_{y}=\int_{y,0}^{1} r_{\gamma(s),y}|\nabla_{s} \gamma|_{y}ds_{y}\\\\\hspace{1cm}=\int_{y,0}^{1} e^{\Theta(\gamma(s))_{y}-\Theta(y)}|\nabla_{s} \gamma|_{y}ds_{y}.\end{array}\end{equation} Here $r_{x,y}(r_{\gamma(s),x})_{y}=r_{\gamma(s),y}$ has been used.

              The description of scaled curve lengths and line elements illustrates a difference between external and internal scaling. For local entities, such as line elements, external scaling can always be removed by using the location of the entity as the reference point. For nonlocal entities, such as curve lengths, the effect of scaling can perhaps be minimized by suitable choice of  a reference point, but it can never be removed. The one exception is the case with no scaling in which $\Theta(x)$ is a constant.

            \subsubsection{Vectors}

            A good illustration of the effect of scaling on curve lengths is that for vectors  in Euclidean space.   If $\gamma$ is a vector from $x$ to $y$ with $\gamma(0)=x$ and $\gamma(1)=y$ then $\gamma(s)=x+sd_{x}\hat{\mu}_{x}$ and $\nabla_{s}\gamma=d_{x}\hat{\mu}$ is independent of $s.$ $\hat{\mu}$ is a unit vector in the direction from $x$ to $y,$ and  $d_{x}=|\nabla_{s}\gamma|_{x}$ is the unscaled length of the vector referred to $\bar{R}_{x}$ at $x.$   Use of the scaling factor in Eq. \ref{LgThx} gives the scaled length of $\gamma$, referred to  the initial point, $x$, of $\gamma$:  \begin{equation}\label{LvectSc1} L(\gamma)^{\Theta}_{x} =d_{x}\int_{0,x}^{1} e^{\Theta(x+sd_{x}\hat{\mu})_{x}-\Theta(x)}ds.\end{equation}

            The length of $\gamma,$ referred to the end point, $y,$  of $\gamma,$ is given by \begin{equation}\label{LvectSc2} L(\gamma)^{\Theta}_{y} =d_{y}\int_{0,y}^{1} e^{\Theta(x+sd_{x}\hat{\mu})_{y}-\Theta(y)}ds. \end{equation} Here $d_{y}$ and $\Theta(x+sd_{x}\hat{\mu})_{y}$ are the same number values in $\bar{R}_{y}$ as $d_{x}$ and $\Theta(x+sd_{x}\hat{\mu})_{x}$ are in $\bar{R}_{x}$.  Comparison of the lengths in these two equations shows that $L(\gamma)^{\Theta}_{y}$ is not the same number value in $\bar{R}_{y}$ as $L(\gamma)^{\Theta}_{x}$ is in $\bar{R}_{x}.$ This follows from parallel transforming $L^{\Theta}(\gamma)_{y}$ to $x$ to obtain \begin{equation}\label{LgVcyx}
            (L^{\Theta}(\gamma)_{y})_{x}=F_{x,y}L^{\Theta}(\gamma)_{y}= d_{x}\int_{x,0}^{1}e^{\Theta(x+sd_{x} \hat{\mu})_{x}-\Theta(y)_{x}}ds.\end{equation} Here $(L^{\Theta}(\gamma)_{y})_{x}$ is the same number value in $\bar{R}_{x}$ as $L^{\Theta}(\gamma)_{y}$ is in $\bar{R}_{y}.$

            Comparison of this result with Eq, \ref{LvectSc1}  shows that $(L^{\Theta}(\gamma)_{y})_{x}\neq L^{\Theta}(\gamma)_{x}.$ This follows from the fact that, in general,   $\Theta(y)_{x}\neq \Theta(x).$ However, if one includes the scaling factor in the transformation $L^{\Theta}(\gamma)_{y}\rightarrow (L^{\Theta}(\gamma)_{y})_{x}$, then the two lengths, referred to $x,$ are the same.  This is shown by \begin{equation}\label{LgVcyx1}\begin{array}{l}(L^{\Theta}(\gamma)_{y})^{\Theta}_{x}= e^{\Theta(y)_{x}- \Theta(x)}F_{x,y}L^{\Theta}(\gamma)_{y}\\\\\hspace{1cm}= d_{x}\int_{x,0}^{1}e^{\Theta(x+sd_{x} \hat{\mu})_{x}- \Theta(x)}ds=L^{\Theta}(\gamma)_{x}.\end{array}\end{equation}

           Eq.  \ref{LvectSc1} shows clearly the effect of the scaling field on vectors. The integral multiplying the unscaled length, $d_{x},$ gives the change in length due to scaling. If $\vec{A}(\gamma(s))=\nabla_{s}\Theta(\gamma(s))$ is parallel to $\nabla_{s}\gamma$, for all $s,$ then $\vec{A}(x+sd_{x}\hat{\mu})\cdot\hat{\mu}=|\vec{A}(x+sd_{x}\hat{\mu})|$ and the exponent is positive for all values of $s.$ In this case the scaling factor is greater than $1.$  Eq. \ref{LvectSc1} shows this effect in that $\Theta(x+sd_{x}\hat{\mu})$  increases from its value at $x$ as $s$ increases.
            If  $\vec{A}(\gamma(u))$ is antiparallel to $\nabla_{u}\gamma$, for all $u,$ then $\Theta(x+sd_{x}\hat{\mu})$ decreases from its value at $x$ as $s$ increases,  and the scaling factor is less than $1.$  If  $\vec{A}(\gamma(s))$ is perpendicular to $\nabla_{s}\gamma$, for all $s,$ then scaling has no effect on the length of the vector as the scaling factor exponent, $\Theta(x+sd_{x} \hat{\mu})_{x}- \Theta(x)=0$ for all  values of $s.$

            \subsection{Dependence of Scaling on Reference Points}
            For reference points outside $Z$ the dependence of scaling on the choice of reference points can be large. For example, the length of a curve from $x$ to $y,$ referred to the beginning at $x,$ can be quite different from the length referred to the end at $y.$  This can be seen by comparison of Eqs. \ref{LgThx} and  \ref{LgThy1}.  These equations give the scaled lengths of a curve from two different reference points, $x$ and $y.$

            The difference in the scaling integrals results from the difference between values of $\Theta(x)$ and $\Theta(y)$.  If the curve, $\gamma,$ is very long, or is in a cosmological region, or is in a region of rapidly varying $\Theta,$ the difference between these two values can be large.

            The lengths $L(\gamma)^{\Theta}_{x}$ and $L(\gamma)^{\Theta}_{y}$ are the lengths that would be calculated by hypothetical observers at $x$ and at $y$, using their respective mathematics based on $\bar{R}_{x}$ and $\bar{R}_{y}.$ The scaling factors in the integrals in Eq. \ref{LgThx}, and Eq. \ref{LgThy1},  represent the scaling of values of the integrand as elements of $\bar{R}_{\gamma(s)}$ for each $s$, to a common point, either  $x$ and $\bar{R}_{x},$  or $y$  and $\bar{R}_{y}.$ Also $L^{\Theta}(\gamma)_{y}$ is a different number value in $\bar{R}_{y}$ than $L^{\Theta}(\gamma)_{x}$ is in $\bar{R}_{x}.$

\begin{figure}[ht]\begin{center}\rotatebox{-90}
              {\resizebox{140pt}{140pt}{\includegraphics[50pt,310pt]
              [420pt,640pt]{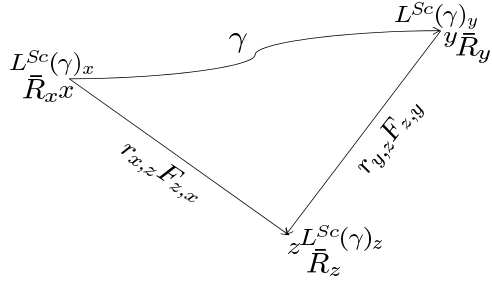}}}\end{center}\caption{The length of curve $\gamma$ for $3$ different locations: the beginning and end of $\gamma$ at $x$ and $y$, and another location, $z,$ away from $\gamma.$  The figure shows that scaled transfer of curve lengths from two different reference points to the same point give a length value that is independent of the original reference points.}\label{ENS4}\end{figure}

            One can parallel transfer, without scaling, the length values,   $L^{\Theta}(\gamma)_{x}$ and  $L^{\Theta}(\gamma)_{y}$, to a common point $z.$  As Eqs. \ref{LvectSc1} and \ref{LvectSc2} show, The transferred values  are different from one another.  If scaling is included, then the transferred values are equal to one another. This follows from  the observation that
            \begin{equation}\label{LgThzyx} e^{\Theta(x)_{z}-\Theta(z)}F_{z,x}L^{\Theta}(\gamma)_{x} =e^{\Theta(y)_{z}-\Theta(z)}F_{z,y}L^{\Theta}(\gamma)_{y}.\end{equation}This is shown schematically in Figure \ref{ENS4}.  

              In a similar way, the scaled line element is affected by a change of reference point. If the reference point for $(dx^{2})^{\Theta}$ is at $x$ then there is no scaling. This follows from Eq. \ref{dx2Sc} with $z$ replaced by $x.$ For other reference points, $y,$ the exponent of the scaling factor is $\Theta (x)_{y}-\Theta(y).$ This clearly depends on the value of $\Theta(y).$ $\Theta(x)_{y}$ is the same  number value in $\bar{R}_{y}$ as $\Theta(x)$ is in $\bar{R}_{x}$. So it is independent of the choice of $y.$

              \subsection{Reference Points in $Z$.}

              The above description of the effect of reference points  applies to  theoretical descriptions of line elements and curve lengths made by observers, $O_{x}$ and $O_{y},$ at $x$ and $y$. For  points, $x$ and $y$ that are cosmological locations, these observers are hypothetical.  However, all mathematical theory descriptions used in physics and geometry are made by  us as observers in $Z.$ The reference points for these descriptions with scaling should be  limited to points in $Z$.

            In Section \ref{RT} it was seen that the effect of scaling was negligible for all points in $Z$.  It follows that the scaling factor for any reference point in $Z$ should be independent of the choice of the point.  It is sufficient to show this for the line element as the results are the same for any other quantity.

            Let $z$ and $w$ be two points in $Z$. The scaled value of $ds^{2}_{x},$ referred to $z$ is given by Eq. \ref{dx2Sc}.  The scaled value referred to $w$ is given by Eq. \ref{dx2Sc} with $w$ replacing $z$ as
           \begin{equation}\label{dx2Scw}
            (dx^{2})^{\Theta}_{w}=e^{\Theta(x)_{w}-\Theta(w)}dx^{2}_{w}.\end{equation} Parallel transformation of this to $z$ for comparison to the scaled line element at $z$ gives \begin{equation}\label{dx2Scwz}
            \begin{array}{l}((dx^{2})^{\Theta}_{w})_{z}=F_{z,w}(dx^{2})^{\Theta}_{w}= e^{\Theta(x)_{z}-\Theta(w)_{z}}dx^{2}_{z}\\\\
            \hspace{1cm}=e^{\Theta(z)-\Theta(w)_{z}}(dx^{2})^{\Theta}_{z}.\end{array}\end{equation} Since $\Theta(z)-\Theta(w)_{z}\simeq 0$ for all $w,z$ in $Z$, one has
            \begin{equation}\label{dx2Sczw}((dx^{2})^{\Theta}_{w})_{z}\simeq (dx^{2})^{\Theta}_{z}.\end{equation}  This proves the independence because $((dx^{2})^{\Theta}_{w})_{z}$ is the same number value in $\bar{R}_{z}$ as $(dx^{2})^{\Theta}_{w}$ is in $\bar{R}_{w}.$

             The same result holds for quantities that include scaling in their definitions.   The length of $\gamma$ at $z$ is related to that at $w$ by  \begin{equation}\label{LgThzw} L^{\Theta}(\gamma)_{z} =e^{\Theta(w)_{z}-\Theta(z)} F_{z,w}L^{\Theta}(\gamma)_{w} \simeq(L^{\Theta}(\gamma)_{w})_{z}.\end{equation} Here $(L^{\Theta}(\gamma)_{w})_{z}$ is the same number value in $\bar{R}_{z}$ as $L^{\Theta}(\gamma)_{w}$ is in $\bar{R}_{w}.$

            The description of the scaled path length, given by Eqs. \ref{LgScx} and \ref{rpathsx}, or by \ref{LgThx}, should apply to many different geometries. Besides Euclidean and Minkowski geometries these equations should apply to Riemann  and Pseudoreimann geometries. For these geometries one would use the metric tensor $g_{\mu,\nu}(x)$ to define the scalar products appearing in the various expressions for the curve length.

             \subsection{Distance Between Points}\label{DBP}
             Since  the length of curves is affected by scaling, one would also expect distances to be affected by scaling. This is the case. The distance between points $x,y$ is defined to be the length of the minimum length curve from $x$ to $y.$

             Let $\gamma$ be a curve where $\gamma(0)=x$ and $\gamma(1)=y.$ The minimum length curve from $x$ to $y$ is found by varying \begin{equation} \label{LgScxTh}L(\gamma)^{\Theta}_{x}=\int_{x,0}^{1}
             e^{\Theta(\gamma(s))_{x}-\Theta(x)}|\nabla_{s} \gamma|_{x}ds_{x}\end{equation}
              with respect to $\gamma$ and setting the result equal to $0.$  Here  $|\nabla_{s}\gamma|=(\nabla_{s}\gamma \cdot\nabla_{s}\gamma)^{1/2}.$
              For each coordinate index value, $\mu,$ the resulting Euler Lagrange equation is \begin{equation} \label{ELEU}\frac{\partial \Theta(\gamma(s))}{\partial\gamma_{\mu}}|\nabla_{s}\gamma|-
             \frac{d}{ds}(\Theta(\gamma(s))) \frac{\partial |\nabla_{s}\gamma|}{\partial (\partial_{\mu,s}\gamma)}=\frac{d}{ds}\frac{\partial |\nabla_{s}\gamma|}{\partial (\partial_{\mu,s}\gamma)}.\end{equation} Here $\partial_{\mu,s}\gamma=d\gamma_{\mu}/ds.$

             The curve, $\gamma,$ satisfying Eq. \ref{ELEU}, is a geodesic or minimum length curve between $x$ and $y.$ The length of $\gamma$ is the distance between $x$ and $y.$

             Eq. \ref{ELEU} shows that the presence of a space and/or time  dependent scalar field, $\Theta,$ adds two terms to the usual Euler Lagrange equation.  If $\Theta(x)$ is a constant for all $x$, then the two left hand terms of Eq. \ref{ELEU} are $0,$ and one obtains the usual equation,
             \begin{equation}\label{ELU} \frac{d}{ds}\frac{\partial |\nabla_{s}\gamma|}{\partial  (\partial_{\mu,s}\gamma)}=0.\end{equation}

             \section{Examples of the Effect of $\Theta$ on Geometry}\label{EETG}
             So far, some general effects of $\Theta$ induced scaling on geometry have been discussed.  It is worth considering some specific examples of the possible space and time dependence of $\Theta$. It is hoped that these will help  to further understand the properties of the $\Theta$ boson  field and its interaction with physical systems and with geometry.

             One way to investigate further the properties and effects of $\Theta$ is to  examine the possible dependence of $\Theta$ on space and time. So far the only requirement is that $\nabla_{z}\Theta\simeq 0$ for all $z$ in $Z$. Outside of $Z$ there are no restrictions.

             \subsection{Time Dependent $\Theta$}\label{TDT}
             An interesting example  is the effect of a possible time dependence of $\Theta$  on geometric and physical quantities.  Assume that $\Theta(x),$ as $\Theta(t,\vec{x})=\Theta(t)$, depends only on the time and not on space. For  $3$ dimensional Euclidean space and nonrelativistic physics, there is no scaling present at any point, $x,$ if the time for the reference point, $z,$ is the same as that for  $x.$

              For example, the scaled line element at $x,$ referenced to a point $z$  for Euclidean space is given by:\begin{equation}\label{dx2Scz}(dx^{2})^{\Theta}_{z} =e^{\Theta(x)_{z}-\Theta(z)} dx^{2}_{z}=e^{\Theta(x)_{z}- \Theta(z)} dx^{j}_{z}dx^{j}_{z}\end{equation} (Sum over the same indices implied). If $x=t,\vec{x}$ and $z=t,\vec{z},$ then $\Theta(x)_{z}-\Theta(z)=\Theta(t)_{z}-\Theta(t)=0,$  and there is no scaling.

              Scaling is present if the reference point time, $t$, is different from the time, $t'$, associated with some event at $\vec{x}.$ This follows from the fact that $\Theta(t')_{z}-\Theta(t)\neq 0$ can occur.

              Scaling is also present in relativistic physics. In order for  any quantity or event at a space time point $x$ to be observable from a reference point, $z,$ it is necessary that $x=t',
              \vec{x}$ be within the past light cone of $z.$   This condition is expressed by \begin{equation}\label{ttpc}t\geq t'+\frac{1}{c}|\vec{x}-\vec{z}|.\end{equation}

              For relativistic physics the scaled line element at $x,$ referenced to $z,$ is given by  \begin{equation}\label{dx2Sct}(dx^{2})^{\Theta}_{z} =e^{\Theta(x)_{z}-\Theta(z)}(\eta_{\mu,\nu}dx^{\mu}dx^{\nu})_{z} =e^{\Theta(x)_{z}- \Theta(z)} [(cdt')^{2}-dx^{j}dx^{j}]_{z}.\end{equation}  The subscript, $z,$  indicates parallel transform from $x$ to $z.$ If $x$ is on the past light cone of $z$ then \begin{equation} \label{tptc} t'=t-\frac{1}{c}|\vec{x}-\vec{z}|\end{equation} and\begin{equation}\label{dx2Sctz} (dx^{2})^{\Theta}_{z} =e^{\Theta(t-(1/c)|\vec{x}-\vec{z}|,\vec{x})_{z} -\Theta(t,\vec{z})} dx^{2}_{z}.\end{equation}

              This equation holds for all reference points $z$ and points $x$ on the past light cone of $z.$ Description of  events at distant locations that are visible to us as observers,  requires that  $z=t,\vec{z}$ is a point in $Z.$  In this case, $t\simeq 14 \times 10^{9}$ years or the age of the universe.

               If $\Theta(t',\vec{x})$ depends only on time and not on $\vec{x},$  the scaling factor in Eq. \ref{dx2Sctz}, \begin{equation}\label{Sctx}
              e^{\Theta(t-(1/c)|\vec{x}-\vec{z}|,\vec{x})_{z} -\Theta(t,\vec{z})}=e^{\Theta(t-(1/c)| \vec{x}-\vec{z}|)_{z} -\Theta(t)},\end{equation} is spherically symmetric about $\vec{z}.$ Its value is unchanged for all $x$ that keep the value of  $|\vec{x}-\vec{z}|$ unchanged.

              As was noted before, the effect of scaling on quantities depends on the gradient of $\Theta$ and not on the value of $\Theta.$  The scaling factor is unaffected by changing the value of $\Theta$ everywhere by a constant. This follows from $(\Theta(x)+c)_{z}-(\Theta(z)+c)=\Theta(x)_{z} -\Theta(z).$

              The result is that one is free to choose the value of $\Theta$ at some location and use that as a refer4ence.  Since the points $z$ in $Z$ are the locations available to us as observers and are the locations from which we attempt to describe the properties of the universe, it is reasonable to use $\Theta(z)$  in $Z$ as a reference. As noted earlier, this value is the same for all $z$ in $Z$. Here the choice, \begin{equation}\label{Tz0}\Theta(z)=0\end{equation}for all $z$ in $Z,$ is made. Use of this value gives  a rewrite of Eq. \ref{dx2Sctz} as \begin{equation}\label{dx2Sct0} (dx^{2})^{\Theta}_{z} =e^{\Theta(t')_{z}}dx^{2}_{z}=e^{\Theta(t-(1/c)|\vec{x}-\vec{z}|)_{z}} dx^{2}_{z}.\end{equation}

              The dependence of the scaling factor on the time, $t',$  with $x$ on the past light cone of $z,$ is of interest.   Two cases are examined here. In one $\Theta(t')$ increases as $t'$ increases from $0$ to $t$, or \begin{equation}\label{dTdtcp}\frac{d\Theta(t')} {dt'}>0.\end{equation} In the other $\Theta(t')$ decreases as $t'$ increases from $0$ to $t$, or \begin{equation}\label{dTdtcn}\frac{d\Theta(t')} {dt'}<0.\end{equation}

                If Eq. \ref{dTdtcp} holds, then  $\Theta(t'),$ and the scaling factor, get larger as $t'$  increases from $0$ to $t,$ or as the spatial distance between $x$ and $z$ approaches $0.$ $\Theta(t')$ reaches a maximum value of $0$ (Eq. \ref{Tz0}) at $t=t'.$ This means that values of scaled physical and geometric entities at times $\leq t$ in the past light cone of $z,$ referenced to $z,$ get smaller with increasing distance between $x$ and $z.$ This includes quantities such as line elements, curve lengths, and distances between points.  In this case $\Theta(t')<0$ and  the scaling factor is less than $1.$

                 Continuing with Eq. \ref{dTdtcp}, it follows that $\Theta(t')$  has a minimum value at $t'=0.$ If $\Theta(t')\rightarrow -\infty$ as $t'\rightarrow 0,$  then the scaling factor approaches $0$. It follows that, at $t'=0,$ or $14\times 10^{9}$ years in the past, numerical values of physical, geometric, and mathematical quantities, describing events at $t',$ are seen by us at time $t,$ as all equal to $0$.  This is reminiscent of the big bang in that  all points of space, and possibly space itself, must be crammed into a point. This follows from the fact that space distances between all point pairs are $0.$ However, it is also the case that  all  other scaled mathematical and physical  quantities, including energy and mass  are also squeezed to $0.$

                There is a problem with this description in that both $\Theta(0)=-\infty$ and $t'=0$ cannot hold.  This is that the value, $0,$ associated with any physical or geometric quantity at $t'=0$ will remain $0$ for all finite $t'.$ The reason is that $0$ is the only number value that is invariant under scaling.

                This can be avoided by either restricting $\Theta$ so that $\Theta(0)$ is a large but finite negative number, or restricting $t'$ to values arbitrarily close but different from $0.$ In this case, values of quantities that are different from $0$ at the present time $t$ will remain different from $0$ for all $t',$ and will scale.

                This description raises the possibility that the scalar $\Theta$ field might be used to describe the expansion of space as is required in models of  inflation \cite{Guth}. An example of this is the use of the scalar inflaton field to describe the expansion of space \cite{Albrecht}.  It is possible for $\Theta(t')$ to increase sufficiently fast from a very large negative value to account for inflation. This follows from the fact that the scaling factor is the exponential of $\Theta(t')$ as in $e^{\Theta(t')},$ and $\Theta(t')$ can increase very fast from a large negative value. Since scaling affects all values in the same way,  energies and other physical quantities increase at the same rate as do distances between points.

                If this possibility has any merit, then  $\Theta(t')$ must  increase very rapidly to give a rapid increase of the scale factor. After a short time when inflation ends, $\dot{\Theta}(t')\equiv d\Theta(t')/dt'$ can decrease to describe a moderate rate of expansion.  If $\dot{\Theta}(t')=k>0$ then $\Theta(t')=k(t'-t)<0$ and the scale factor is $e^{k(t'-t)}$.  If $k$ is such that $kt'<1,$ then expansion to first order gives $e^{k(t'-t)}\simeq e^{-kt}(1+kt')=1+kt'$ ($e^{-kt}=1$, Eq. \ref{Tz0}), and the expansion rate is constant.  Physical and geometric quantities, such as distances between points, increase at a constant rate, $k$. If $\dot{\Theta}(t')=kt'$, then the scale factor, $e^{k((t')^{2}-t^{2}/2},$ would show that distances between points, and possibly space itself, are expanding at an accelerating rate.

                If the accelerated increase of distances does apply to space,  then the effect of $\Theta$ is similar to that of dark energy \cite{Li,Bousso} as far as the accelerated expansion of space is concerned. If quintessence \cite{Ost,Zlatev,Steinhardt} is the scalar field for dark energy, then $\Theta$ acts in a similar fashion to quintessence as far as space expansion is concerned.

                  If eq. \ref{dTdtcn} holds, then the scaling factor increases as $t'$ decreases.  If $d\Theta(t')/dt'=-k$ with $k>0,$ then the scaling factor at time $t'$ is $e^{k(t-t')}.$ For values of $t'$ where $kt'<1,$ the scaling factor, $e^{k(t-t')}\simeq e^{kt}(1-kt')=1-kt',$ is linear in $t'.$  This corresponds to the contraction of distances and, possibly, of space as $t'$ increases.

\begin{figure}[h]\begin{center}\vspace{1.7cm}\rotatebox{-90}
              {\resizebox{140pt}{140pt}{\includegraphics[180pt,300pt]
              [510pt,630pt]{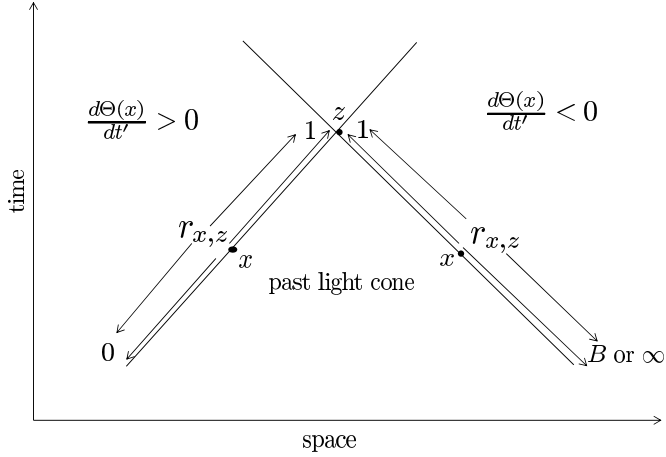}}}\end{center}\caption{Effect of time dependence of $\Theta(x)$ on scaling. The left and right hand sides of the past light cone are used for the two cases of $d\Theta(x)/dt' >0$ and $d\Theta(x)/dt' <0.$ $r_{x,z}=e^{\Theta(x)}$ is the scaling factor. The arrows show the effect on scaling as $x=t',\vec{x}$ moves toward and away from $z=t,\vec{z}$ along the light cone. $B$ denotes an upper bound in case the scaling factor is bounded from above.}\label{ENS5}\end{figure}

                Figure \ref{ENS5} illustrates these two cases of positive and negative time derivatives as $x$ moves towards and away from $z$ along the past light cone of $z.$ To save on space, one light cone is used for these two cases. $r_{x,z}$ is the scaling factor.

              It has been seen that the space and time dependence of scaled mathematical quantities that depend on numbers depends on the space time dependence of the boson field, $\Theta.$ This raises the possibility that numbers may be treated as physical systems, \cite{Szekeley}.  Whether this idea has merit or not is a question for the future.

             For general relativity the equation expressing the time dependence of the value of the scaled line element, Eq. \ref{dx2Scz} with $\Theta(z)=0,$ becomes  \begin{equation}\label{dx2SczGR}(dx^{2})^{\Theta}_{z} =e^{\Theta(x)_{z}} dx^{2}_{z}=e^{\Theta(x)_{z}} g_{\mu,\nu}(x)_{z}dx^{\mu}_{z}dx^{\nu}_{z}.\end{equation}
            The FRW line element \cite{FRW} provides a good example for scaling.  The usual expression for the line element is \begin{equation}\label{FRW}dx^{2}=c^{2}dt^{2}-a(t)^{2}(\frac{dr^{2}}{1-kr^{2}} +r^{2}d\Omega^{2}).\end{equation}  With separate mathematical universes at each point of $M$, this represents the line element at $x.$ The unscaled representation at any other point $z$ is obtained by parallel transporting the terms to $z.$ This gives \begin{equation}\label{FRWz}dx^{2}_{z}=(c^{2}dt^{2})_{z}-a(t)^{2}_{z}(\frac{dr^{2}}{1-kr^{2}} +r^{2}d\Omega^{2})_{z}.\end{equation} Each term has the same  value  in the mathematics at $z$ as the corresponding term in Eq. \ref{FRW} has in the mathematics at $x.$

            The scaled representation of the FRW line element is given by \begin{equation}\label{FRWzSc} \begin{array}{l}(dx^{2})^{\Theta}_{z}=(e^{\Theta(x)_{z}})\{(c^{2}dt^{2})_{z} -a(t)^{2}_{z}(\frac{dr^{2}}{1-kr^{2}}+r^{2}d\Omega^{2})_{z}\}\\\\\hspace{1cm}= (e^{\Theta(x)_{z}})(c^{2}dt^{2})_{z}-(e^{\Theta(x)_{z}})a(t)^{2}_{z} (\frac{dr^{2}}{1-kr^{2}} +r^{2}d\Omega^{2})_{z}.\end{array}\end{equation}  This equation shows that both the time component $(c^{2}dt^{2})_{z} $ and the time dependent space component, $a(t)^{2}_{z} (\frac{dr^{2}}{1-kr^{2}} +r^{2}d\Omega^{2})_{z}$ are multiplied by the same scaling factor.  It also shows that the time dependence of the factor $a(t),$ which is based on the Einstein equations \cite{FRW}, is not related to the time dependence of $\Theta$ and scaling  as described here.

             It should be emphasized again that $(dx^{2})^{\Theta}_{z}$ is $O_{z}^{\prime}s$ description, at $z,$ of the line element at $x$. The scaling factors relate the number values at $x$ to those at $z$. This is the case for all number values at $x.$ The scaling factors are the same for all numbers.  They are independent of what physical quantities, if any, that are associated with the numbers.

             In spite of this caveat, it is also the case that, for locations $x$ that are cosmological for observers in $Z$, $O_{z}$ must use scaled descriptions of events and aspects of space and time that are very far away from $Z.$ The reason is that $O_{z}'s$ mathematics is based on  number and other mathematical systems that are local to $z.$ Mathematical descriptions of events or properties at distant points are  based on mathematics that is local to these points. Since these  are not available to observers in $Z,$ the descriptions and predictions must be transferred to the mathematics local to $O_{z}$.  These transferrals use both parallel transformations and scaling.

             \subsection{Black and White Scaling Holes}\label{BWSH}
             So far descriptions of points $x$ at which the boson field, $\Theta(x),$ varies rapidly have been limited to times, $t'\approx 0,$ or close to the time of the big bang. However there is no reason why $\Theta(x)$ cannot vary rapidly and assume  very large positive or negative values at other points. Just as gravitational singularities in space lead to black holes, singularities in values of the field, $\Theta(x),$ can lead to both black and white scaling holes.

             Two examples will be described here: one for black scaling holes and the other for white scaling holes. Both are cases in which the absolute value of $\Theta(x)$ approaches infinity as $x$ approaches a point $x_{0}$ which is far away from $Z.$ For black scaling holes, $\Theta(x)\rightarrow \infty$ as $x\rightarrow x_{0}.$ For white scaling holes, $\Theta(x)\rightarrow -\infty$ as $x\rightarrow x_{0}.$ To keep things simple, $\Theta(x)$ is assumed to depend on space only and not on time.

\begin{figure}[ht]\begin{center}\vspace{1cm}\rotatebox{-90}
              {\resizebox{140pt}{140pt}{\includegraphics[100pt,300pt]
              [430pt,630pt]{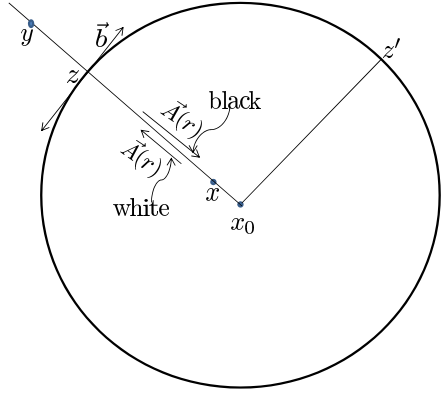}}}\end{center}\caption{Illustration for  black and white scaling holes. The directions of $\vec{A}(r)=\nabla_{r}\Theta,$ which is the gradient of $\Theta$ at $r,$ are shown for these two cases. $z$ is an arbitrary reference point far away from $Z$. The circle is the locus of constant values of $\Theta$ at a radial distance, $l=|z-x_{0}|$ from $x_{0}.$ Here $r$ is any radial distance from $x_{0}.$ The figure shows that  $\vec{A}(z)\cdot\vec{b}=0$ where $\vec{b}$ is tangent to the circle at $z.$ }\label{ENS6}\end{figure}

              Let the field $\Theta$ be spherically symmetric about $x_{0}$.  Then it depends solely on the radial distance, $r,$ from $x_{0}$ to $x.$ The radial gradient of $\Theta$,  $\vec{A}(r),$ is either parallel or antiparallel to the radius vector from $x_{0}.$  Figure \ref{ENS6} illustrates the setup.

              Let $\gamma$ be a vector from $z$ to $x_{0}$ parameterized by $0\leq s\leq 1$ where $\gamma(s) =z +sl\hat{\mu}.$ Here $\hat{\mu}$ is a unit vector along the radius from $z$ to $x_{0},$  and $l$ is the unscaled length of $\gamma.$  The scaled length to points $x$ along $\gamma$,   referred to $z,$ is given by Eq. \ref{LvectSc1}  as \begin{equation}\label{LvectT} L(\gamma)^{\Theta}_{z} =l\int_{0,z}^{w} e^{\Theta(z+sl\hat{\mu})_{z}-\Theta(z)}ds.\end{equation} Here $x=z+wl\hat{\mu}.$ Also $0\leq w\leq 1.$

              Consider  the case where $\vec{A}(\gamma(s))=\nabla_{s}\gamma$ is directed towards $x_{0}$ along a radius as in Figure \ref{ENS6}. Assume that $\Theta$ varies with distance along a radius according to \begin{equation}\label{blkT}\Theta(\gamma(s))=\frac{K}{|x_{0}-\gamma(s)|} =\frac{K}{l(1-s)}.\end{equation} Here $|x_{0}-\gamma(s)|$ is the distance from $x_{0}$ to $\gamma(s)$. Also $K>0.$ For this example $\Theta(\gamma(s))$ has a positive singularity at $s=1.$

              Use of Eq. \ref{blkT} in Eq. \ref{LvectT} gives \begin{equation}\label{LvectT1} L(\gamma)^{\Theta}(w)_{z} =l\int_{0,z}^{w} e^{\Theta(\gamma(s))_{z}-\Theta(\gamma(0))}ds =l\int_{0,z}^{w} \exp(\frac{K}{l-sl}-\frac{K}{l})ds.\end{equation}This integral can be simplified by arbitrarily choosing $K$ and $z$ so that $K/l=1.$

              This allows simplification of the integral to \begin{equation}\label{LScrznd}L^{\Theta}(w)_{z}= l\int_{z,0}^{w}\exp(\frac{1}{1-s}-1)ds.\end{equation}   The dimensionless integral is multiplied by $l$ or $K$ to give the integral the dimensions of length.

\begin{figure}[h]\begin{center}\vspace{2cm}\rotatebox{-90}
              {\resizebox{140pt}{180pt}{\includegraphics[180pt,240pt]
              [510pt,570pt]{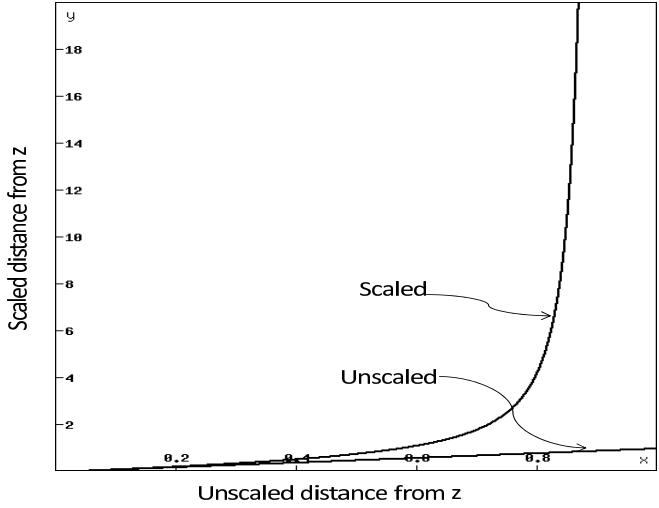}}}\end{center}\caption{Scaled and unscaled dimensionless distances from
              a reference point $z$ to $x_{0}$ where $\Theta(x_{0})=\infty,$  $\Theta(x)
              =K/|x_{0}-x|,$ and $K>0.$ Distances are shown as a function of $w$ where $0\leq w\leq 1.$ }\label{ENS7}\end{figure}

              Figure \ref{ENS7} is a plot  of the integral in Eq. \ref{LScrznd} as a function of $w.$ As such it shows the scaled  and unscaled dimensionless distances from $z$ to points, $x,$  along the radius to $x_{0}.$  The abscissa  gives the  unscaled distance from $z$ to $x$
              as $w$ in  dimensionless units,  and the ordinate gives the scaled distance, also in dimensionless units.

              Figure \ref{ENS7} shows dramatically that the scaled distance from $z$ to $x$ goes to infinity as $x$ approaches $x_{0}$, or $w\rightarrow 1.$   For instance, for $x$ where the distance from $z$ to $x$  is $80\%$ of that from $z$ to $x_{0},$ the scaled distance is about $4$ times that of the unscaled distance.

\begin{figure}[h!]\begin{center}\vspace{2.3cm}\rotatebox{-90}
              {\resizebox{140pt}{180pt}{\includegraphics[190pt,240pt]
              [520pt,570pt]{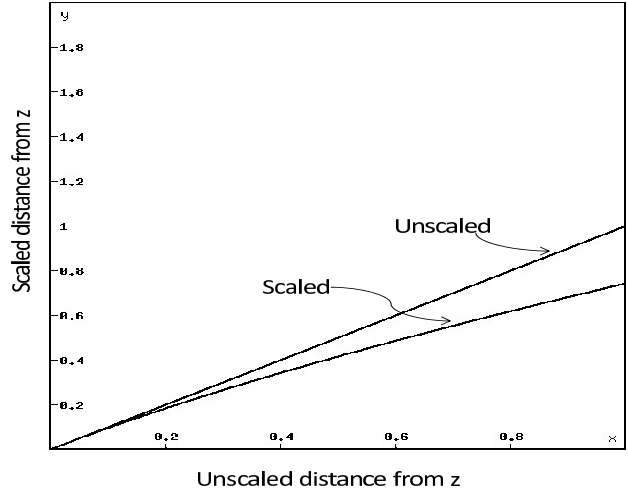}}}\end{center}\caption{Scaled and unscaled dimensionless distances from $x_{0}$ to points from $z$ to $z'.$ Values are referenced to $z$ for comparison with those in the previous figure. $z$ and $z'$ are arbitrarily chosen so that $|x_{0}-z|/K=1$ and $|x_{0}-z'|/K=2.$}\label{ENS8}\end{figure}

              Figure \ref{ENS8} is a plot of the scaled and unscaled dimensionless distances from $z$ to a point, $z'$, (Fig. \ref{ENS6}) more distant from $x_{0}$ than $z$. The scaled distance, referenced to $z$, is a plot of the integral in \begin{equation}\label{LTzzp}L^{\Theta}(w)_{z} =l'\int_{0,z}^{w} \exp(\frac{1}{s+1}-1)ds \end{equation} as a function of $w$ with $0\leq w\leq 1.$ The path length between $z$ and $z'$ is denoted by $l'.$ $z'$ is chosen so that $|x_{0}-z'|/K=2.$ The unscaled distances from $z$, as a straight line, are also shown for comparison.

              The integral is obtained by noting that path length from $x_{0}$ extends from $|z-x_{0}|$ to $|z'-x_{0}|,$ or from $1$ to $2$ in dimensionless units. The scaled path is shorter than the unscaled path because  the location of the reference point, $z,$ is closer to $x_{0}$ than is any point between $z$ and $z'.$ Also the direction of $\vec{A}(r)$ towards $x_{0}$ is opposite to the path direction. This compresses the path length.

              This setup, with the scaling vector $\nabla_{r}\Theta=\vec{A}(r)$ directed towards  positive infinity at $x_{0},$ is denoted here as a \emph{scaling black hole}. One reason is that the scaled distances from a reference point,  $z,$ to radial points, $x,$ increase without bound as $x\rightarrow x_{0}.$  Another reason is that if one puts a classical system at $z$ with nonzero energy, then the force acting on the system is towards $x_{0}.$ This is shown in Eq. \ref{ScEL}, which is the equation of motion obtained from the action with scaling included.

              The other case of interest here is obtained by setting $K<0$ in Eq. \ref{blkT}.  In this case $\Theta(x)\rightarrow -\infty$ as $x\rightarrow x_{0}$, As shown in Fig. \ref{ENS6}, $\vec{A}(x)$ is directed radially outward from $x_{0}$ with  magnitude becoming infinite as $x$ approaches $x_{0}.$

              For this case the scaled distance from $z$ to $x$, referenced to $z,$ is given in dimensionless units by Eq. \ref{LScrznd} with  the exponent, $1/(1-s)-1,$ of the integrand replaced by  its negative, $1/(s-1)+1.$ As before, $z$ is such that $|x_{0}-z|/|K|=l/|K|=1.$  The integral upper limit remains as $w.$ $|K|$ is the absolute value of $K$.

\begin{figure}[h!]\begin{center}\vspace{2.4cm}\rotatebox{-90}
              {\resizebox{140pt}{180pt}{\includegraphics[180pt,240pt]
              [510pt,570pt]{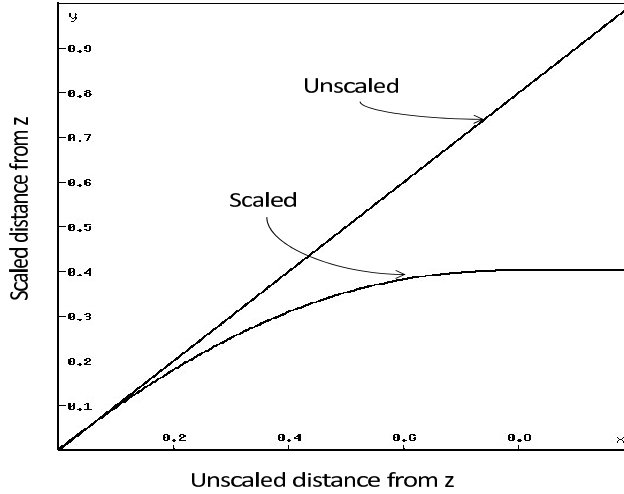}}}\end{center}\caption{Scaled and unscaled dimensionless distances from $z$ to points on the radius from $z$ to $x_{0}.$ Values are referenced to $z.$ $z$ is arbitrarily chosen so that $l/|K|=1.$ The direction of the potential gradient  of $\Theta$ is opposite to that of the path from $z$ to $x_{0}.$}\label{ENS9}\end{figure}

              The results are shown in Figure \ref{ENS9} as a plot of the integral as a function of $w$ where, as before, $0\leq w\leq 1.$ 
              The curves in this figure show that the scaled distance from $z$ to $x$, Fig. \ref{ENS6}, approaches a limit at about $0.4$ of the unscaled distance. It is close to this limit for values of  $x$ that are greater than $70\%$ of the way from $z$ to $x_{0}.$

              In a sense, the outward directed vector field $\vec{A}(r)$ acts like a barrier preventing the scaled distance from reaching its unscaled value. If one puts a classical system at $z$, then as Eq. \ref{ScEL} shows, the direction of the force acting on the particle is radially away from $x_{0}.$  For these reasons, this example of $\Theta$ is called a \emph{scaling white hole} at $x_{0}.$

              \section{Discussion}\label{D}

              There are several aspects of the effect of the boson field, $\Theta,$ on physics and geometry  that should be noted. One aspect is the appearance  and emphasis on the notion of "sameness".  This shows up in gauge theories where the unitary operator, $U_{y,x},$  in Eq. \ref{UZW}, defines or describes the meaning of "same vector" between two points.  This is different from the usual treatment of gauge theories where the notion of "same vector"  does not appear if one  represents $U_{y,x}$ using generators of a Lie algebra. However, as was noted, this is not correct mathematically.

              The concept of "sameness" also plays an important role in the basic model used here with separate  mathematical structures associated with different space time points.  The parallel transformation operators, $F_{y,x},$  Eq. \ref{Fyx}, for numbers and unitary operators, $U_{y,x},$  Eq. \ref{UZW}, for vector spaces describe or correspond to this concept. If $a_{x}$ is a number value in the complex number structure, $\bar{C}_{x}$, then $a_{y}=F_{y,x}a_{x}$ is the same number value in $\bar{C}_{y}$ as $a_{x}$ is in $\bar{C}_{x}.$ If $\psi_{x}$ is a vector in the vector space, $\bar{V}_{x},$ then $\psi_{y}=U_{y,x}\psi_{x}$ is the same vector in $\bar{V}_{y}$ as $\psi_{x}$ is in $\bar{V}_{x}.$

              As far as the definitions of parallel transforms \cite{Mack} are concerned, it makes no difference here whether the notion of "sameness" is assumed a priori and the operators $F_{y,x}$ and $U_{y,x}$ are required to preserve it, or the choice of the operators defines the notion of  "sameness" between mathematical structures.  Here the concept of sameness or same value is taken for granted, and it makes no difference whether "sameness" or parallel transformation operators are taken to be a priori.

              The concept of "sameness" or "same value"  provides a reference point or base for the description  and meaning of  scaling between structures at different points.   As was noted earlier in this work, the model assumption of separate mathematical structures at each space time point, with parallel transform operators between structures, and no scaling, gives the same mathematics and physical theory predictions as does the usual mathematical setup. This shows that separate mathematical structures at each point, with an appropriate notion of sameness and no scaling, is completely equivalent to  the usual treatment of mathematics and its use in physics.  The possibility that one might be able to dispense with the concept of sameness or parallel transformation and treat these and scaling together as emergent concepts is an open question.

              Probably the main question to consider is what physical field, if any, does $\Theta$ represent.  The appearance of $\Theta$ as a scalar boson field with very weak coupling to matter fields in gauge theory helps, but it does not answer this question. The fact that there is no evidence of the presence of $\Theta$ on physics in a local region, $Z$, Section \ref{RT}, restricts the effects of $\Theta$ to cosmological regions.  If $\Theta$ has any observable  or predictable consequences, they would show up over very large regions of space and time.

              It is interesting to note that there are many suggestions in the literature for the  role that  scalar fields  have in physics \cite{Linder}. Many are attempts to explain dark energy \cite{Li,Bousso}.  Some  include modification of gravity by including a scalar field \cite{Brans,Jordan}. Others use scalar fields to describe a space and time varying  cosmological constant, $\Lambda.$   Quintessence  is one example that has been much studied \cite{Ost,Zlatev,Steinhardt}. Other examples of  scalar fields in physics include the  inflaton to describe inflation in cosmology \cite{Albrecht}, the Higgs boson \cite{Higgs}, and supersymmetric partners of spin $1/2$ fermions \cite{Martin}.

              At present it is not known if the  field, $\Theta,$ described in this work, corresponds to any of these physical fields.  These examples do show that there are many possibilities. It is also possible that $\Theta$ combines mathematics and physics together in a manner that is different from that of corresponding to any of the existing fields.

              One potentially desirable approach would be to consider the maps, with scaling between two number structures as single maps instead of their description as  products of two maps, a parallel transformation followed by a scaling.  In this approach, the concept of "same number as" and the representation of the map as a product of two maps would be an emergent property.

              It is clear from these considerations that there is much to do in order to tie the model considered here more closely to physics, if it is possible to do so. In addition, more needs to be done to determine if physics makes use of the $\Theta$ field. In any case one
              should keep in mind that, for the model considered here with separate mathematical structures at each space and time point,   the scaling factors and parallel transformations provide mappings between these mathematical structures and their elements.  They apply to numbers and other mathematical entities independent of whether the entities represent physical quantities or have  nothing to do with physics. However, as has been seen, their effect on physical quantities, especially those represented by derivatives or integrals over space and time, can be appreciable.

              \section{Conclusion}

              In this work the assignment of separate vector spaces to each space time point in gauge theories was expanded to assign separate complex number structures to each point. The freedom of choice of bases in the vector spaces was expanded to include freedom of choice of scaling for the complex scalars. Use of this with the requirement that Lagrangians be invariant under local gauge transformations, resulted in the presence of a boson field, $\Theta,$ whose mass is optional and interaction with matter fields is very weak.

              The second part of the paper explores a basic model that  extends the idea of separate complex number structures at each point to include structures for other types of numbers and mathematical systems that are based on numbers. The mathematics available to an observer, $O_{x}$, at $x,$ is assumed to be limited to the structures at $x.$ In other words, mathematics is local. Maps between the structures are products of parallel transforms and  $\Theta$ induced scaling factors.

              The effects of $\Theta$ scaling on some aspects of physics and geometry are investigated.  The fact that experiments in physics give no hint of scaling implies that scaling must be absent for all points $z$ in a region $Z$ occupiable by us as observers. For all $x,y$ in $Z,$ $\Theta(y)-\Theta(x)\simeq 0.$ Outside $Z$ at cosmological distances, this restriction does not apply.

              Mathematical descriptions of events or properties at far away locations, $x$,  use the mathematics at $x$.  The description by $O_{z}$, using the mathematics at $z,$ requires transfer with scaling of the description at $x$ to $z.$ This type of scaling is external in that the transfer is of the mathematical description. If the mathematical description at $x$ includes space or time integrals or derivatives, then there is a scaling factor inside the integral or derivative. This type of scaling is internal. Internal scaling differs from external scaling in that external scaling can be removed by transferring the reference point to the location $x.$  Internal scaling cannot be so removed.

              The effects of scaling on geometric properties of entities outside $Z$ was  investigated. In particular, line elements and  lengths of curves were described both with and without the effects of scaling. It was seen that with separate mathematical structures at each point but without scaling, the description, using parallel transform operators, gave the same results as the usual model with just one global set of mathematical structures. The scaling factors based on $\Theta$ were also found to be independent of  geometry in the sense that  they appear as  multiplying factors that are independent of the metric tensor.

              The effect of the choice of reference points on scaling was described. One good feature is that the effect of scaling on physics and geometry is  the same for all reference points in $Z$.   All observers in $Z$  should agree on the presence or absence of $\Theta$ induced scaling,  and the effect should be the same for all observers.

              Two specific examples of the effect of   $\Theta$ on geometry are investigated.  In one, $\Theta$ is assumed to be time dependent and independent of space location,
              $\Theta(t',\vec{x})=\Theta(t').$ If $x$ is on the past light cone of any point $z$ in $Z$ and $d\Theta(t')/dt'>0$, then distances,  line elements, and other quantities increase as $t'$ increases.  $t'=0$ is the time of the big bang. Since $\Theta(t')$ can increase very rapidly from a very large negative value at $t'$ equal or close to $0$, scaling can describe the cramming of space into a small volume or point. It can also describe the rapid increase of space like that described by  inflation \cite{Guth}.

              At later times, if $d\Theta(t')/dt'=k>0,$ or $kt'$, the time dependence of $\Theta$ can describe the constant, or accelerated expansion of distances between points. In this sense it mimics dark energy \cite{Li,Bousso}. If $d\Theta(t')/dt'<0$, then the time dependence of $\Theta$ describes the contraction of distances and other quantities as $t'$ increases from $0$ to  the present time, $t\approx 14\times 10^{9}$ years.

              The  other example shows that $\Theta$ can give rise to what are denoted here as black and white scaling holes. Black scaling holes arise in case $\Theta(r,\theta,\phi)=\Theta(r)$ is spherically symmetric around some point $x_{0}$ and $\Theta(r)\rightarrow\infty$ as $r\rightarrow 0.$ Here $r$ is the radial distance between some point and $x_{0}.$ A specific example where $\Theta(r)=K/r$  and $K>0$ is worked out.

              This is denoted a black scaling hole because the path length from a point $z$ to any point $x$ on the radius from $z$ to $x_{0}$ increases without bound as $x$ approaches $x_{0}.$ Also if a particle is placed at point $z$, then one finds that the force on the particle due to $\Theta$ equals the product of $\vec{A}(r)$ and the Lagrangian density. As such it is directed towards $x_{0}$ and  increases without bound.

              White scaling holes arise in case $\Theta(r)\rightarrow -\infty$ as $r\rightarrow 0.$ A specific example with $\Theta(r)=K/r$  and $K<0$ is described. In this case  scaled path lengths between $z$ and $x$ approach a barrier in that they approach a maximum of about $40\%$ of the unscaled path length between $z$ and $x_{0}$ as $x$ approaches $x_{0}.$ The force on a particle at $x$ is directed outward along the radius. It also increases without bound as $x$ approaches $x_{0}.$

              As was noted in the discussion, there is much to be done. If, and it is a big if, one can show that physics and geometry make use of the boson field $\Theta$, either as one of the scalar fields already proposed, or to tie mathematics and physics more closely together, then real progress will have been made towards constructing a coherent theory \cite{BenCTPM1,BenCTPM2} of physics and mathematics together.

             \section*{Acknowledgement}
            This work was supported by the U.S. Department of Energy,
            Office of Nuclear Physics, under Contract No.
            DE-AC02-06CH11357.

             \end{document}